\documentstyle[12pt,aasms4,graphics]{article}
\newcommand{\etal}{et~al.}
\def\baselinestretch{1.4}
\begin{document}
\oddsidemargin=0mm

\title{
SN 2002cx: The Most Peculiar Known Type Ia Supernova
}

\author{Weidong Li\altaffilmark{1}, Alexei V. Filippenko\altaffilmark{1},
Ryan Chornock\altaffilmark{1}, Edo Berger\altaffilmark{2}, Perry
Berlind\altaffilmark{3},
Michael L. Calkins\altaffilmark{3}, 
Peter Challis\altaffilmark{3}, 
Chris Fassnacht\altaffilmark{4,5}, Saurabh Jha\altaffilmark{1,3}, 
Robert P. Kirshner\altaffilmark{3}, 
Thomas Matheson\altaffilmark{3}, Wallace L. W. Sargent\altaffilmark{2}, 
Robert A. Simcoe\altaffilmark{2}, Graeme H. Smith\altaffilmark{6}, and Gordon 
Squires\altaffilmark{7} \\
Email: (wli, alex, rchornock)@astro.berkeley.edu}

\altaffiltext{1}{Department of Astronomy, University of California, Berkeley,
CA 94720-3411.}

\altaffiltext{2}{Department of Astronomy, 105-24 Caltech, Pasadena,
CA 91125.}

\altaffiltext{3}{Harvard/Smithsonian Center for Astrophysics, 60 Garden St.,
Cambridge, MA 02138.}

\altaffiltext{4}{Space Telescope Science Institute, 3700 San Martin Dr.,
Baltimore, MD 21218.}

\altaffiltext{5}{Department of Physics, University of California, Davis,
CA 95616.}

\altaffiltext{6}{UCO/Lick Observatory, University
of California, Santa Cruz, CA 95064.}

\altaffiltext{7}{SIRTF Science Center, 220-6 Caltech, Pasadena, CA 91125.}

\slugcomment{Submitted to PASP}

\begin{abstract}

We present photometric and spectroscopic observations of supernova (SN) 2002cx,
which reveal it to be unique among all observed type Ia supernovae (SNe~Ia). SN
2002cx exhibits a SN 1991T-like premaximum spectrum, a SN 1991bg-like
luminosity, and expansion velocities roughly half those of normal SNe Ia.
Photometrically, SN 2002cx has a broad peak in the $R$ band and a plateau phase
in the $I$ band, and slow late-time decline. The $(B - V)$ color evolution is
nearly normal, but the $(V - R)$ and $(V - I)$ colors are very red.  Early-time
spectra of SN 2002cx evolve very quickly and are dominated by lines from
Fe-group elements; features from intermediate-mass elements (Ca, S, Si) are
weak or absent.  Mysterious emission lines are observed around 7000~\AA\ at
about 3 weeks after maximum brightness. The nebular spectrum of SN 2002cx is
also unique, consisting of narrow iron and cobalt lines. The observations of SN
2002cx are inconsistent with the observed spectral/photometric sequence, and
provide a major challenge to our understanding of SNe~Ia.  No existing
theoretical model can successfully explain all observed aspects of SN 2002cx.

\end{abstract}


\section{INTRODUCTION}

Observations of Type Ia supernovae (SNe~Ia) in the last decade have
substantially advanced our understanding of them. Within the current
observational scheme of SNe~Ia, the majority are the so-called ``normal" or
``Branch normal" variety (Branch, Fisher, \& Nugent 1993), while the others are
``peculiar" SNe Ia which can be further divided into SN 1991T-like or SN
1991bg-like objects (see Filippenko 1997, and references therein). Li et
al. (2001a) discuss SN 1999aa-like objects as another potential subclass of the
peculiar SNe~Ia.

Spectroscopically, normal SNe~Ia before or near maximum light show conspicuous
absorption features near 6150~\AA\ due to Si~II, and near 3750~\AA\ due to
Ca~II, as well as absorption features attributed to Co~II, S~II, and O~I. SN
1991T-like objects (e.g., Filippenko et al. 1992b) before maximum light show
unusually weak lines of Si~II, S~II, and Ca~II, yet prominent high-excitation
features of Fe~III. The usual Si~II, S~II, and Ca~II lines develop in the
postmaximum spectra, and by a few weeks past maximum the spectra look nearly
normal. SN 1999aa-like objects (Li et al. 2001a) are similar to the SN
1991T-like ones, but with significant Ca~II H \& K absorption lines in their
premaximum spectra. SN 1991bg-like objects (e.g., Filippenko et al. 1992a) near
maximum light show a broad absorption trough extending from about 4100 to
4400~\AA\, due to Ti~II lines, and enhanced Si~II/Ti~II $\lambda$5800
absorption relative to Si~II $\lambda$6150.

Photometrically, normal SNe~Ia show a correlation between the peak luminosity
and light-curve decline rate. This was first convincingly demonstrated by
Phillips (1993), and subsequently exploited by Hamuy et al.  (1996a), Riess,
Press, \& Kirshner (1996), Perlmutter et al. (1997), and Phillips et
al. (1999). The slower, broader light curves are intrinsically brighter at peak
than the faster, narrower light curves. This trend seems to extend to the
peculiar events: SN 1991T-like objects display a slow light-curve evolution
compared to normal SNe~Ia, and are thought to be more luminous than average,
while SN 1991bg-like objects show a fast light-curve evolution, as well as
intrinsically red colors at maximum light.  The latter are also about 2 mag
fainter in the $B$ band than normal SNe~Ia.  In addition, Nugent et al. (1995)
demonstrate that the relative strengths of some spectral features correlate
with the peak luminosity of SNe~Ia, and that the gross spectral variations
among all SNe~Ia can be accounted for by simply varying the photospheric
temperature, suggesting that peculiar events are just the extreme tails of a
continuous distribution.

At first glance, it seems that almost {\it all} SNe~Ia form a sequence that can
be described by a single parameter.  One such parameter is the decline rate
[e.g., $\Delta m_{15}(B)$, the decline in magnitudes during the first 15 days
after the maximum brightness in the $B$ band]. Another such parameter is
$R$(Si~II), the ratio of the depths of two absorption features near 5800 and
6150~\AA\, that are often attributed to Si~II $\lambda$5972 and Si~II
$\lambda$6355, respectively (Nugent et al. 1995; note that Garnavich et
al. 2002 show that the 5800~\AA\ feature is actually mostly Ti~II in
low-luminosity SNe~Ia). A one-parameter description of SNe~Ia is useful, and in
the context of Chandrasekhar-mass explosions it can be interpreted in terms of
a variation in the mass of ejected radioactive $^{56}$Ni.

However, theoretical explosion models of SNe~Ia (e.g., H\"oflich \& Khokhlov
1996, hereafter HK96; H\"oflich et al. 1996) suggest that not all SNe~Ia can be
put on a one-parameter sequence. Observationally, Hamuy et al. (1996b) show
that some light curves with similar decline rates have significant differences
in particular details. Also, some SNe~Ia with normal-looking spectra (i.e.,
having only lines of the usual ions) exhibit exceptionally high blueshifts of
their absorption features (Branch 1987), and Wells et al. (1994) find no
correlation between the blueshift of the Si~II $\lambda$6355 absorption near
the time of maximum light and the decline rate in a small sample of
well-observed SNe~Ia (see also Patat et al. 1996). Hatano et al. (2000) also
demonstrate that the spectroscopic diversity among SNe~Ia is multi-dimensional.

Perhaps the most disturbing published case for the one-parameter description of
SNe~Ia is SN 2000cx (Li et al. 2001b). Photometrically, SN 2000cx differs from
all known SNe~Ia, and its light curves cannot be fit well by the fitting
techniques currently available. There is an apparent asymmetry in the $B$-band
peak, in which the premaximum brightening is relatively fast (similar to that
of the normal SN 1994D), but the postmaximum decline is relatively slow
(similar to that of the overluminous SN 1991T).  The color evolution of SN
2000cx is also peculiar: the $(B - V)_0$ color has a unique plateau phase and
the $(V - R)_0$ and $(V - I)_0$ colors are very blue. SN 2000cx also has a
unique spectral evolution.  The premaximum spectra of SN 2000cx are similar to
those of SN 1991T-like objects, having weak Si~II and prominent Fe~III
lines. The Si~II lines emerge near maximum light and stay strong until about
three weeks past maximum. The Fe~III lines also remain prominent until well
after maximum, implying a rather slow change in the excitation stages of the
iron-peak elements.  The expansion velocities, as derived from the absorption
lines of the iron-peak and intermediate-mass element lines, are unusually high.

In this paper we present observations of another bizarre object, SN~Ia 2002cx,
many of whose properties are the exact opposite of those of SN 2000cx.  SN
2002cx was discovered by Wood-Vasey et al. (2002) with the Oschin 1.2-m
telescope at Palomar Observatory in unfiltered images taken on 2002 May 12.2
and May 16.2 UT (UT dates are used throughout this paper). Wood-Vasey et al.
(2002) measured a precise position for SN 2002cx as $\alpha$ =
13$^h$13$^m$49$^s$.72, $\delta = +6\arcdeg 57\arcmin 31\farcs 9$ (equinox
J2000.0), which is 18$\arcsec$ south and 11$\arcsec$ east of the center of its
host galaxy CGCG 044-035.  An optical spectrum obtained with the Fred Lawrence
Whipple Observatory (FLWO) 1.5-m telescope on May 17.2 (Matheson et al. 2002)
identified the object as a peculiar SN 1991T-like event at about a week before
maximum brightness, but the supernova is {\it underluminous} (instead of
overluminous) compared with normal SNe~Ia.  The Si~II $\lambda$6355 and Ca~II H
\& K lines are extremely weak or absent, but the Fe~III lines at 4300~\AA\ and
5000~\AA\, are present.

  We recognized the uniqueness of SN 2002cx (SN 1991T-like premaximum spectrum,
SN 1991bg-like luminosity, and very low expansion velocity) shortly after its
discovery, and a follow-up program of multicolor photometry was established at
Lick Observatory.  Spectra of the SN were obtained with the FLWO 1.5-m telescope
and also with the Keck 10-m telescopes.  This paper presents the results from
this campaign and is organized as follows. Section 2 contains a description of
the observations and analysis of the photometry, including our methods of
performing photometry, our calibration of the measurements onto the standard
Johnson-Cousins system, our resulting multicolor light curves, and our
comparisons between the light curves and color curves of SN 2002cx and those of
other SNe~Ia. Section 3 contains a description of the spectral observations and
analysis. We discuss the implications of our observations in \S 4 and summarize
our conclusions in \S 5.

\section{PHOTOMETRY}

\subsection{Observations and Data Reduction}

Broadband $BVRI$ images of SN 2002cx were obtained using an Apogee AP7 CCD
camera with the 0.76-m Katzman Automatic Imaging Telescope (KAIT; Li et
al. 2000; Filippenko et al. 2001) at Lick Observatory. The Apogee camera has a
back-illuminated SITe 512$\times$512 pixel CCD chip. At the $f/8.2$ Cassegrain
focus of KAIT, the 24 $\mu$m pixel of the chip yields a scale of 0$\farcs$8
pixel$^{-1}$, making the total field of view of the camera 6$\arcmin.7\times
6\arcmin.7$. The typical seeing at KAIT is around 3$\arcsec$ full width at half
maximum (FWHM), so the CCD images are well sampled.

$BVRI$ images of SN 2002cx were also obtained using dewar No. 2 with the 1.0-m
Nickel telescope at Lick Observatory. The camera with the back-illuminated
2048$\times$2048 pixel Loral chip was used in a 2$\times$2 binned mode,
yielding a scale of 0$\farcs$36 pixel$^{-1}$ and a total field of view of
6$\arcmin.1 \times 6\arcmin.1$. The typical images have FWHM $\approx$
2$\arcsec$ and thus are also well sampled.

Figure 1 shows a KAIT $V$-band image taken on 2002 May 18 with SN 2002cx and
eight local standard stars marked. Absolute calibration of the field was done
with KAIT on 2002 June 6 and 11, and with the Nickel telescope on 2002 May 17,
June 10, 11, and 12 by observing Landolt (1992) standard stars at different
airmasses throughout the photometric nights. Instrumental magnitudes for the
standard stars were measured using aperture photometry with the
IRAF\footnote{IRAF (Image Reduction and Analysis Facility) is distributed by
the National Optical Astronomy Observatories, which are operated by the
Association of Universities for Research in Astronomy, Inc., under cooperative
agreement with the National Science Foundation.}  DAOPHOT package (Stetson
1987) and then employed to determine transformation coefficients to the standard
Johnson-Cousins $BVRI$ system. The derived transformation coefficients and
color terms were then used to calibrate the sequence of eight local standard
stars in the SN 2002cx field. The magnitudes of these eight stars and the
associated uncertainties derived by averaging over the photometric nights are
listed in Table 1.  Notice that the local standard stars have different numbers
of calibrations (column 6 in Table 1) because the two telescopes have different
total fields of view.

We tried the point-spread-function (PSF) fitting method (Stetson 1987) to
perform differential photometry of SN 2002cx relative to the comparison stars,
but the results are less than satisfactory.  As can be seen in Figure 1, SN
2002cx is contaminated by its host galaxy (especially in the $R$ and $I$
bands), and the relatively poor resolution of KAIT images together with seeing
variations yield fluctuations at the $\pm 0.1$--0.2 mag level in the final
light curves (lower panel in Figure 2). The PSF fitting method also
overestimates the brightness of SN 2002cx, as a negative residual can be seen
at the position of the SN on the processed images with SN 2002cx and the
comparison stars subtracted.

The solution for getting precise photometry of SN 2002cx is to obtain $BVRI$
template images after the SN fades, and apply galaxy subtraction to remove the
galaxy contamination. We have attempted to get these template images with both
KAIT and the Nickel telescope (which has better resolution than the KAIT data)
when SN 2002cx was 7 months old, but unfortunately, owing to its slow late-time
decline (see below for details), it is still faintly visible in these images.
Since it will take considerable time for SN 2002cx to fade sufficiently, we
have instead employed a pseudo-galaxy-subtraction technique. We chose one
Nickel image that has the best seeing (FWHM $\approx$
1$\arcsec$.2--1$\arcsec$.5; obtained on 10 July 2002, $\sim 2$ months past
maximum light) in each passband, performed PSF-fitting photometry, and
subtracted only the SN from the image.  This SN-subtracted image was then used
as a template to perform galaxy subtraction on all the remaining KAIT and
Nickel images in the same passband.

We have fine-tuned the parameters in PSF fitting so that these ``poor-man's
templates" have minimal residuals at the position of the SN.  Nevertheless, a
very faint negative residual still remains in these ``template"
images. Although our results therefore systematically overestimate the
brightness of SN 2002cx, we consider them to be superior to those measured from
the PSF-fitting technique for the following reasons. (1) The
galaxy-subtraction technique removes the galaxy contamination, so seeing
variations do not significantly affect the photometry. Consequently, the final
light curves are much smoother than those from the PSF-fitting technique. (2)
As can be seen from the lower panel of Figure 2, the PSF-fitting photometry is
usually about 0.1--0.3 mag brighter than the galaxy-subtraction photometry. As
the galaxy-subtraction photometry already overestimates the brightness of the
SN, the PSF-fitting overestimates the brightness even more. In nights of good
seeing (the 1st, 5th, and 6th Nickel observations), both techniques give
similar measurements. 

For the $B$ and $V$ bands, we estimate that our galaxy-subtraction technique
overestimates the brightness of SN 2002cx by $\sim 0.03$ mag around maximum
and $\sim 0.2$ mag at late times. For the $R$ and $I$ bands, the overestimate 
is larger, perhaps $\sim 0.1$ mag around maximum and 0.3 mag at late times. 
In late 2003 or 2004, when SN 2002cx becomes very faint, we plan to obtain
genuine template images and derive more accurate, final magnitudes.

The PSF-fitting technique is used on the galaxy-subtracted images to measure
the instrumental magnitudes for the SN and the comparison stars, which are then
used to obtain the standard Johnson-Cousins $BVRI$ magnitudes of SN 2002cx by
doing differential photometry.  The color terms involved in this transformation
are $-$0.04, 0.04, 0.07, and $-$0.01 for the KAIT $B, V, R,$ and $I$ filters,
respectively, while they are $-$0.08, 0.06, 0.10, and $-$0.04 for the Nickel
$B, V, R,$ and $I$ filters, respectively. Our preliminary $BVRI$ measurements
of SN 2000cx are listed in Table 2.  Uncertainties for the measurements were
estimated by combining in quadrature the errors given by the photometry
routines in DAOPHOT with those introduced by the transformation of instrumental
magnitudes onto the standard system. The systematic uncertainty caused by the
galaxy subtraction is not considered, and is $\sim 0.1$--0.3 mag as discussed
above.

\subsection{Optical Light Curves}

The upper panel of Figure 2 displays our preliminary $BVRI$ light curves of SN
2002cx; final light curves will be derived with genuine templates later. The
KAIT data points are shown with open circles and the Nickel ones with solid
circles. For most of the points the statistical uncertainties are smaller than
the plotted symbols.  The overall agreement between the measurements from the
two telescopes is excellent. Hence, in all subsequent discussions and figures,
the KAIT and the Nickel datasets are combined and discussed together. Also
defined here is the variable $t$, which is the time since maximum brightness in
the $B$ band (JD = 2452415.2; see discussion below).

The date and the magnitude of the peak in each passband are listed in 
Table 3. For the $B$ and $V$ bands, these were determined by fitting a
second-order spline function to the data points around maximum
brightness; for the $R$ and $I$ bands, there is no well-defined 
peak and only the magnitude on the broad plateau phase is measured
as the peak. Also listed are the $\Delta m_{15}(B)$ and $\Delta m_{15}(V)$
values.

At first glance, some peculiarities of SN 2002cx can be easily discerned from
Figure 2. SN 2002cx does not have the typical $R$-band and $I$-band secondary
maxima as expected for a SN 1991T-like object; instead, it has a broad peak
(or plateau phase) not seen in any other SNe~Ia, followed by a slow decline at
late times. To facilitate comparisons, we plot in Figure 3 the light curves of
SN 2002cx and those of several other well-observed SNe~Ia representing the
diversity of SN Ia light curves: SN 1991T [$\Delta m_{15}(B)=0.95\pm0.05$; Lira
et al. 1998] as an example of an overluminous SN Ia; SN 1994D [$\Delta
m_{15}(B) =1.31\pm0.08$; Richmond et al. 1995] as an example of a normal SN Ia;
and SN 1991bg [$\Delta m_{15}(B)=1.93 \pm 0.08$; Filippenko et al. 1992a;
Leibundgut et al. 1993] as an example of a subluminous SN Ia.  Also included in
the comparison are the light curves of SNe 2000cx and 1999ac.  SN 2000cx
[$\Delta m_{15}(B)=0.93 \pm 0.04$; Li et al. 2001b] is the opposite of SN
2002cx in many respects, while SN 1999ac [$\Delta m_{15}(B)=1.30 \pm 0.09$; Li
et al. in preparation] is the closest match to SN 2002cx we can find in the
literature and in our unpublished photometric database. All light curves are
shifted in time and peak magnitude to match those of SN 2002cx with the time
zero-point being the date of maximum light in the $B$ band.

The $B$-band light curve of SN 2002cx (Fig. 3a) before $t$ = 15~d shows a very
similar evolution to that of SN 1999ac: their premaximum brightening is faster
than SN 1991T, but slower than SNe 1994D and 2000cx. Their postmaximum decline
is faster than SNe 1991T and 2000cx, and similar to SN 1994D. In fact, SNe
1994D, 1999ac, and 2002cx have very similar $\Delta m_{15}(B)$
values\footnote{Test reductions using the galaxy-subtraction technique with
template images obtained when SN 2002cx was 7 months old suggests that $\Delta
m_{15}(B)$ of SN 2002cx may be around 1.6 mag, higher than the 1.3 mag reported
here. However, the contamination of the SN light in these particular template
images is unclear. }. As $\Delta m_{15}(B)$ is often considered to be a good
indicator of the luminosity of a SN Ia (Phillips 1993; Hamuy et al. 1996a;
Phillips et al. 1999), the similar decline rates of the three SNe implies that
they should have similar absolute magnitudes. As discussed in \S 4.2, however,
this is not the case: SN 2002cx is less luminous than the other two SNe by as
much as 1.4 mag in the $B$ band.

After $t$ = 15~d, SN 1999ac deviates from SN 1994D and evolves more slowly. SN 2002cx
shows the same trend but is more extreme. The data point at $t$ = 50~d, though
uncertain, suggests that SN 2002cx has the smallest decline among
all SNe in the comparison during the first 50 days after the $B$-band maximum
brightness.

The $V$-band light curve of SN 2002cx (Fig. 3b) shows a similar evolution to
that of SN 1999ac until $t$ = 30~d: the premaximum phase of SN 2002cx is not
well observed and seems to be consistent with SN 1999ac, SN 1994D, or SN
2000cx.  The postmaximum decline of both SN 2002cx and SN 1999ac is faster than
SN 1991T, but slower than the other SNe in the comparison. After $t$ = 30~d, SN
2002cx shows a dramatic change in its decline rate, and has the smallest
late-time decline rate among all the SNe in the comparison.

The $R$-band light curve of SN 2002cx (Fig. 3c) shows a
peculiar evolution. The unfiltered magnitudes reported by Wood-Vasey et
al. (2002) are plotted here as two open triangles with an arbitrary error bar of
$\pm0.2$ mag. These unfiltered magnitudes and the early KAIT data points show
that SN 2002cx has the fastest premaximum brightening among all the SNe in the
comparison.  SN 2002cx also shows quite different evolution than SN 1999ac in
the $R$ band, despite their similarities in the early $B$ and $V$ light curves.
SN 2002cx does not have a ``shoulder" or secondary maximum, but rather a very
broad peak extending to 2 weeks after the $B$-band maximum
brightness. The subluminous SN 1991bg does not have a secondary maximum either,
but it has a much narrower peak around maximum. SN 2002cx also has the slowest
late-time decline rate among all the SNe in the comparison.

The $I$-band light curve of SN 2002cx (Fig. 3d) is also
peculiar. The two earliest data points suggest that SN 2002cx has a rather
steep rise to maximum; it then stays at a nearly constant magnitude until $t$ =
20~d (a plateau phase), after which it begins a slow late-time decline phase.
Jha (2002) shows that this plateau-like behavior, lasting for about 20 d, could
be an extrapolation of a trend seen in $I$-band light curves of subluminous
SNe~Ia, but the subsequent slow decline remains puzzling.

It is also clear from Figure 3 that the photometric behavior of SN 2002cx and
SN 2000cx are often the opposites of each other (e.g., the late-time decline
rates, and the evolution around the $R$-band and $I$-band secondary maxima).

In summary, the optical light curves of SN 2002cx differ from those of all
known SNe~Ia.  SN 1999ac seems to bear some similarities to SN 2002cx in the
early $B$ and $V$ light curves, but their $R$ and $I$ light curves and
late-time $B$ and $V$ evolution are quite different.  The broad peak in the $R$
band and the plateau phase in the $I$ band are puzzling, as are the slow
late-time decline.  More discussion of the absolute magnitude of SN 2002cx, the
$R$-band and $I$-band evolution around the secondary maximum, and the late-time
decline rate can be found in \S 4.

\subsection{Optical Color Curves}

There are indications that SN 2002cx suffers from little host-galaxy
reddening. The premaximum spectra, as discussed in \S 3.2, are quite blue, an
indication of low reddening toward the SN. Moreover, the medium-resolution Keck
ESI spectrum of SN 2002cx taken on June 10 showed negligible interstellar
Na~I~D absorption lines at the redshift of the host galaxy.  In subsequent
discussions, we assume no host-galaxy extinction and use the Galactic reddening
of $E(B - V) =0.034$ mag (Schlegel, Finkbeiner, \& Davis 1998) as the adopted
total reddening for SN 2002cx.

In Figure 4 we present the comparison between the intrinsic optical color
curves of SN 2002cx [$(B - V)_0$, $(V - R)_0$, and $(V - I)_0$] and those of
several other SNe Ia (1991T, 1991bg, 1994D, 1999ac, and 2000cx). The reddenings
adopted for the comparison SNe are $E(B - V)$ = 0.13 mag for SN 1991T (Phillips
et al. 1992), 0.05 mag for SN 1991bg (Filippenko et al. 1992a), 0.04 mag for SN
1994D (Richmond et al. 1995), 0.05 mag for SN 1999ac (Li et al. in
preparation), and 0.08 mag for SN 2000cx (Li et al. 2001b).

The $(B - V)_0$ color curve of SN 2002cx is very similar to that of SN
1999ac. For $t <$ 20~d, this is a natural result as the two SNe have very
similar $B$ and $V$ light curves.  It is somewhat surprising that the late-time
point of SN 2002cx at $t$ = 50~d is also consistent with SN 1999ac, because
the two SNe have different late-time $B$ and $V$ evolution.  However,
this data point is quite uncertain, and the slow decline of SN 2002cx in both
the $B$ and $V$ bands may conspire to yield a color similar to that of SN
1999ac.

SN 2002cx has $(B - V)_0 = -0.04 \pm 0.04$ mag at $t = -4$~d,
and  $(B - V)_0 = 0.04 \pm 0.05$ mag at the time of the $B$ maximum.
These colors are similar to the other SNe in the comparison except 
SNe 1991bg and 2000cx, both of which are redder. Although SN 2002cx
and SN 1991bg have similar absolute magnitudes (see details in \S 4.2), 
SN 2002cx is much bluer than SN 1991bg at $B$ maximum [$(B - V)_0 = 0.04$ mag
for SN 2002cx versus $(B - V)_0 = 0.69$ mag for SN 1991bg]. This significant
difference in color between the two SNe will be used to constrain 
their models in \S 4.7. 

Between $t$ = 5 and $t$ = 20~d, the $(B - V)_0$ color curves of SN 2002cx and
1999ac are redder than those of the other SNe in the comparison except SN
1991bg.  The late-time $(B - V)_0$ colors of SN 1999ac and SN 2002cx are
consistent with the ``Lira-Phillips law" (Lira 1995; Phillips et al.  1999),
though we caution that the late-time $(B - V)_0$ color of SN 2002cx is not well
observed.

The $(V - R)_0$ color evolution of SN 2002cx is peculiar. Before $t$ = 5~d, the
color of SN 2002cx evolves in a manner similar to that of the other SNe except
SN 1991bg, though it may be somewhat redder. Between $t$ = 5 and $t$ = 17~d,
the color of SN 2002cx becomes progressively redder, and it has the second
reddest color among all the SNe in the comparison (only SN 1991bg is
redder). SN 2002cx does not have the dip at $t \approx 12$~d shown by SNe
1991T, 1994D, and 2000cx.  SN 1999ac seems to follow a similar trend, but is
less extreme. After $t$ = 17~d, SN 2002cx becomes redder until $t$ = 25~d, and
then maintains nearly the same red color at later times; SN 1991bg, on other
hand, becomes progressively bluer at later times. Consequently, SN 2002cx
has the reddest color among all the SNe in the comparison after $t$ =
20~d. With the exception of SN 1991bg, SN 2002cx and SN 2000cx seem to be at
opposite extremes: SN 2002cx has the reddest color while SN 2000cx has the
bluest.

SN 2002cx has a red $(V - I)_0$ color at all times: it has the second-reddest
color before $t$ = 25~d (only SN 1991bg is redder), and the reddest color at
later times. SN 2002cx does not have the dip at $t \approx 10$~d exhibited by
the other SNe (except SN 1991bg).  Although SN 1999ac showed some similarities
to SN 2002cx (e.g., the early $B$ and $V$ light curves, the $(B - V)_0$ color
curves), it has a very different $(V - I)_0$ color evolution than SN 2002cx
(but similar to that of SNe 1991T and 1994D). The $(V - I)_0$ color evolution
of SN 2002cx and SN 2000cx also clearly demonstrate that they are opposites:
among all the SNe in the comparison (with the exception of SN 1991bg), SN
2002cx has the reddest color while SN 2000cx has the bluest.

In summary, SN 2002cx has a nearly normal $(B - V)_0$ evolution, and is quite
blue near maximum, but it has red $(V - R)_0$ and $(V - I)_0$ evolution
compared to the other SNe~Ia. In \S 3, we will investigate how the
peculiarities in the light curves and color curves of SN 2002cx are correlated
with its spectral evolution.

\section{SPECTROSCOPY}

\subsection{General Results} 

Optical spectra of SN 2002cx were obtained with the FLWO 1.5-m telescope using
the FAST spectrograph, and with the Keck 10-m telescopes using the Low
Resolution Imaging Spectrometer (LRIS; Oke et al. 1995) and the Echellette
Spectrograph and Imager (ESI; Sheinis et al. 2002).  The journal of
observations is given in Table 4.

All one-dimensional sky-subtracted spectra were extracted optimally in the
usual manner. Each spectrum was wavelength and flux calibrated, and corrected
for continuum atmospheric extinction and telluric absorption bands (Bessell
1999; Matheson 2000). In general, the position angle of the slit was aligned
along the parallactic angle for the observations, so that the spectral shape
does not suffer from differential light loss (Filippenko 1982). The LRIS
spectrum obtained on day 27 had second-order contamination redward of 7800~\AA,
so only the unaffected part was used; it was combined with the day 26 spectrum
to improve the signal-to-noise ratio.

The spectral evolution of SN 2002cx from $t$ = $-$4 to 56~d is shown in Figure
5. All spectra shown in this paper have been corrected for the redshifts of the
host galaxies. For SN 2002cx, a redshift of 7184 km s$^{-1}$ was adopted from
NED\footnote{NED (NASA/IPAC Extragalactic Database) is operated by the Jet
Propulsion Laboratory, California Institute of Technology, under contract with
the National Aeronautics and Space Administration.}. A nebular spectrum of SN
2002cx obtained in January 2003 with Keck using LRIS at $t \approx 240$~d will be
reported elsewhere.

Our strategy for studying the spectral evolution of SN 2002cx is to use Figure
5 as a guide, and to conduct detailed comparisons between SN 2002cx and other
SNe Ia at different epochs in Figures 6 through 9.  We have used SN 1994D
(Patat et al. 1996; Filippenko 1997) as an example of a normal SN Ia, SN 1997br
as an overluminous SN 1991T-like object (Li et al.  1999), and SN 1991bg as a
subluminous object (Filippenko et al. 1992a).  Also included in the comparison
are SN 2000cx (Li et al. 2001b) and SN 1999ac (Li et al. in preparation).
Depending on the availability of the observations for each SN, not all SNe are
shown in each figure. The line identifications adopted here are taken from
Kirshner et al. (1993), Jeffery et al. (1992), Mazzali, Danziger, \& Turatto
(1995), and Mazzali et al. (1997).

\subsection{The Premaximum Phase}

The first spectrum of SN 2002cx was obtained with the FLWO 1.5-m telescope on
2002 May 17, 4 days before the $B$-band maximum brightness. Figure 6 shows a
comparison of this spectrum with those of other SNe Ia at similar epochs.  The
spectrum of SN 2002cx is similar to that of SN 1997br: both show a blue
continuum, with Fe~III $\lambda$4404 and Fe~III $\lambda$5129 being the two
major absorption lines. The small wiggles between 4000~\AA\ and 5000~\AA\ all
match up quite well.  A weak Si~II $\lambda$6355 line with an asymmetric
profile is already present in the spectra of SNe 1997br and 2000cx, but it is
not apparent in the spectrum of SN 2002cx. SN 2002cx does not have strong Ca~II
H \& K lines, which suggests that it belongs to the subclass of genuine SN
1991T-like objects, and not to the subclass of SN 1999aa-like objects
discussed by Li et al. (2001a).

However, the spectra of SN 2002cx and SN 1997br show a significant difference
in the expansion velocities derived (using the special relativistic formula)
from the absorption minima of the Fe~III lines. As qualitatively indicated by
the dotted vertical lines in Figure 6, the Fe~III absorption minima of SN
2002cx are at much redder wavelengths than SN 1997br. Quantitatively, the
expansion velocities measured from the Fe~III lines are $\sim$ 10,400 km
s$^{-1}$ for SN 1997br, while they are only $\sim$ 6,400 km s$^{-1}$ for SN
2002cx, the lowest expansion velocity ever measured for a SN~Ia at such early
times. As more fully discussed in \S~3.6 and \S~4.5, low expansion velocity is
a distinct characteristic of SN 2002cx that places strong constraints on the
theoretical models and provides clues to its peculiar photometric and spectral
evolution.

Mazzali et al. (1995) modeled the early-time spectrum of SN 1991T and suggested
that the outer part of the ejecta is dominated by Fe-group elements, which give
rise to the Fe~III lines in the spectra.  They also show that the absence of
the Si~II and Ca~II lines is due mainly to the high envelope temperature, which
is caused by the high luminosity of SN 1991T as a direct consequence of
overproduction of $^{56}$Ni.  At times around and after maximum brightness, the
abundance of the Fe-group elements drops relative to that of the intermediate
mass elements (IME). Consequently, a weak but apparent Si~II $\lambda$6355 line
appears in the $t = -3$~d spectrum of SN 1991T (Phillips et al. 1992) and in
similar events like SN 1997br (Li et al.  1999).

SN 2002cx, however, may have a different premaximum spectral formation
mechanism than that of SNe 1991T and 1997br.  While the Fe~III lines are likely
caused by the presence of Fe-group materials in the outer part of the ejecta,
the high temperature of the envelope may be caused by a small ejecta mass
rather than the overproduction of $^{56}$Ni. As discussed in \S 4.2, SN 2002cx
is $\sim 2$ mag subluminous compared with normal SNe~Ia at all optical bands, so
it has a rather small $^{56}$Ni yield, probably of order $0.1~M_\odot$.
However, this small quantity of $^{56}$Ni is distributed in the outer part of
the ejecta and heats the small envelope to high temperatures. More arguments
for a small ejecta mass in SN 2002cx are discussed in \S 4.4.

There is also evidence that part of the reason the Si~II and Ca~II lines are
absent in the premaximum spectra of SN 2002cx is the small amount of IME in the
ejecta. As discussed later, little IME may have been synthesized in the
explosion, a distinct characteristic of SN 2002cx.

The FLWO spectrum of SN 2002cx at $t = -1$~d showed very little evolution,
compared with the $t = -4$~d spectrum (Fig. 5), although the Fe~III absorption
lines are somewhat deeper.

\subsection{Two Weeks after Maximum}

Our next two spectra of SN 2002cx were obtained with the FLWO 1.5-m telescope on
June 2 and 6, 12 and 16 days after $B$ maximum, respectively (Fig. 5). The two
spectra show only subtle differences, so we have used the $t$ = 12~d spectrum
as the representative of the two. The spectrum has undergone a
significant change since $t$ = $-$1~d: the continuum is much redder, the Fe~III
lines are gone, and Fe~II lines are prominent.  We tried to compare the $t$ = 12~d
spectrum to that of other SNe~Ia at similar epochs, but could not find a good
match. A better comparison is achieved when the spectrum is plotted along with
other SNe~Ia at considerably older ages, such as 3 weeks after maximum
(Fig. 7).

Unlike other SNe~Ia such as SNe 1997br, 1994D, and 1991bg, SN 2002cx has
particularly weak Ca~II H \& K lines. Other IME features are also weak: no
Si~II $\lambda$6355 line can be unambiguously identified, and the Si~II/Ti~II
plus Na~I~D line at 5700~\AA\ is weaker than in the other SNe~Ia. There is no
clear evidence for Ti~II lines at around 4100--4400 \AA\, as showed by the
subluminous SN 1991bg (Filippenko et al. 1992a). Fe~II lines, on the other
hand, are quite strong: Fe~II $\lambda$4555 and Fe~II $\lambda$5129 form two
absorption troughs at the dotted vertical lines in Figure 7, and the Fe~II
lines at the blue and red wings of Si~II $\lambda$6355 are much stronger than
in the other SNe. The closest match to the day 12 spectrum of SN 2002cx is that
of SN 1997br at $t$ = 21~d (they only differ significantly in the Ca~II H \& K
lines, besides the different expansion velocities). As discussed by Li et
al. (1999), SN 1997br also has an earlier transition than SN 1991T to strong
Fe~II lines.

The extremely fast spectral evolution of SN 2002cx between $t$ = $-$1~d and
12~d is in dramatic contrast to its earlier and later evolution: the spectrum
of SN 2002cx does not evolve much between $t$ = $-$4~d and $-$1~d as discussed
above, nor does it between $t$ = 12~d and 27~d as seen in Figure 5: the three
spectra taken on days 20, 25, 26/27 are almost identical, and even the day 12
and 16 spectra show similar lines, though with a bluer continuum and relatively
``washed out" features. The extremely fast spectral evolution of SN 2002cx
during the two weeks after maximum brightness suggests that SN 2002cx has very
low-mass ejecta. Its photospheric phase is short, and thus it enters the
nebular phase quickly.

We also note the apparent differences between SN 2002cx and SN 2000cx (Li et
al. 2001b): SN 2000cx has a very slow transition from Fe~III to Fe~II lines,
persistent Si~II lines, and high expansion velocities, while SN 2002cx has
exactly the opposite characteristics.

\subsection{Three Weeks after Maximum}

The Keck spectra of SN 2002cx at $t$ = 20~d (taken with ESI), and at $t$ = 25 and
26/27~d (taken with LRIS), exhibit only subtle differences.  Accordingly, we have
combined the day 20, 25, and 26/27 spectra to construct a spectrum with a range from
3400~\AA\ to 10000~\AA\ and compared it to the spectra of other SNe~Ia in Figure
8. Again, a good comparison can only be achieved when this spectrum of SN
2002cx is plotted against other SNe~Ia at older ages. The spectrum is still
dominated by strong Fe~II lines. This is the first time the spectrum redward of
7400~\AA\ was observed for SN 2002cx, and it shows a much weaker Ca~II infrared
(IR) triplet compared with the other SNe~Ia, consistent with the Ca~II H \& K line
comparison. The O~I $\lambda$7773 line was also observed, with strength
comparable to the other SNe~Ia. All lines are measured to have much smaller
expansion velocities than in the other SNe~Ia.

The superb signal-to-noise ratio of the combined Keck spectrum reveals
additional peculiarities of SN 2002cx. First, whereas the other SNe show broad
emissions, SN 2002cx seems to have the lines split; the emission lines are
``double-peaked." The pairs of vertical lines above the SN 2002cx spectrum
indicate apparent double peaks. There are two possible explanations for the
cause of these double peaks: (1) they come from a common physical origin, such
as the emission from a jet-like configuration, or from rotating clumpy ejecta;
or (2) they are simply lines being resolved in SN 2002cx due to its unusually
low expansion velocity.  However, there is a fatal problem with the first
hypothesis: the separations of the two peaks should have the same (or similar)
velocity difference if they all come from the same physical origin, so the
separation in \AA\ units should increase with increasing wavelength. As can be
seen in the upper panel of Figure 8, the separations do not follow this trend
and are instead quite randomly distributed. Moreover, the lower panel of Figure
8 shows the $t$ = 25~d spectrum of SN 2002cx convolved with a Gaussian function
that has $\sigma$ = 2500 km s$^{-1}$; it looks very similar to SN 1997br at $t$
= 47~d, and all the double peaks are gone.

As can be seen even in the convolved spectrum, however, there are indications
that SN 2002cx may be affected by additional lines (e.g., the region around
7000~\AA).  In the upper panel of Figure 8, the short vertical lines followed
by a ``?" below the SN 2002cx spectrum indicate possible additional lines.
Some are undoubtedly lines being resolved in SN 2002cx, but the three
absorptions and their companion emissions between 6400~\AA\ and 7000~\AA\ do
not have any counterparts in the other SNe~Ia in the comparison.  Figure 9
shows the evolution of this region in greater detail. The three dashed vertical
lines mark the approximate central wavelength of the emission component of
these mysterious lines.  These features are likely caused by additional
emissions rather than absorptions, as the SN enters its emission-dominated
nebular phase early. The 6600~\AA\ line seems to be present even in the day 12
spectrum. These lines have comparable strength and do not evolve much between
$t$ = 20 and 27~d, but they all disappear (or become much weaker) in the $t$ =
56~d spectrum. One might speculate that the 6600~\AA\ line is H$\alpha$ due to
their wavelength coincidence, but the lack of other hydrogen Balmer lines in
the spectra does not support this identification. Moreover, the similar
evolution of these three emission lines strongly suggests that they have
the same origin.

In an attempt to identify these lines, we examined the results of Hatano et
al. (1999a), who presented optical spectra for 45 individual ions that are
candidates for producing identifiable spectral features in SNe. We also
consulted a web-based program, ``the Atomic Line List v2.04,''\footnote{URL
http://www.pa.uky.edu/$\sim$peter/atomic/} where more than 920,000 spectral
transitions are compiled.  We found several ions having transitions with
wavelengths around the marked emissions in Figure 9: C~I, Ca~I], Fe~II, N~II,
Ni~II, and Ti~II.  Of these, we prefer the Ni~II and/or Fe~II identification,
as there are strong Ni~II and Fe~II lines in other parts of the spectrum. When
the SN becomes more nebular, conditions in the ejecta may favor [Ni~II] and
[Fe~II] lines; thus, the three lines disappear in the spectrum at $t$ =
56~d. This suggestion, however, does not explain why the Ni~II and Fe~II
features in the other part of the spectrum do not evolve much between $t$ =
26/27 and 56~d.  We conclude that detailed spectral modeling is required to
reveal the true identity of these lines.

We also note that the $t$ = 43~d spectrum of SN 1999ac shows some signs of the
``double peaks," and some low-amplitude wiggles near 6200~\AA\ and 7000~\AA,
but the overall spectrum looks much more similar to SN 2000cx than to SN
2002cx.

\subsection{Two Months after Maximum}

Our last spectrum of SN 2002cx was taken on July 16, 56~d after $B$
maximum. Figure 10 shows the comparison of this spectrum with that of other
SNe~Ia at much older ages. The spectrum is clearly in the nebular phase, with
emission dominating over absorption. Compared with the other SNe, the features
in the spectrum of SN 2002cx are considerably narrower and appear ``washed out"
--- that is, there are no strong emission or absorption lines. The Ca~II IR
triplet is the strongest line, but it is still much weaker than in the other
SNe.

The overall spectrum of SN 2002cx is quite different from that of the other
SNe~Ia.  The short vertical lines above the spectrum of SN 2002cx mark some of
the differences between SN 2002cx and SNe 2000cx, 1994D, and 1999ac, all of
which have similar nebular spectra. The most significant differences are in the
region between 6500~\AA\ and 8500~\AA, where SN 2002cx has a flat continuum
with weak emissions and weak Ca~II IR triplet absorption, while the other three
SNe have a strong absorption around 6700~\AA, a broad peak near 7300~\AA, and
strong Ca~II IR triplet absorption. There are also apparent differences near
5500~\AA.

The short vertical lines below the SN 2002cx spectrum mark some of the
differences between SN 2002cx and SN 1991bg. Again, the differences are most
significant in the region between 6500~\AA\ and 8500~\AA, where SN 1991bg shows
a broad absorption around 6700~\AA, and a strong emission at 7300~\AA. The
Ca~II IR triplet absorption in SN 1991bg is only slightly stronger than in SN
2002cx, and has a sharp blue wing similar to that of SN 2002cx. The blue wing
of the Ca~II IR triplet of SNe 2000cx, 1994D, and 1999ac, on the other hand,
does not have a sharp edge, and extends to much higher expansion
velocities. This suggests that the calcium is probably confined to a stratified
region in SN 2002cx and SN 1991bg, and it extends to higher velocities in SNe
2000cx, 1994D, and 1999ac.  Another possibility is that there is additional
absorption blueward of the Ca~II IR triplet in SNe 2000cx, 1994D, and 1999ac.

Figure 11 shows possible identifications of the emission lines in the $t$ =
56~d spectrum of SN 2002cx. Whenever possible, the identifications are adopted
from the NLTE spectrum synthesis for the nebular spectra of SN 1991bg (Mazzali
et al. 1997). Because SN 2002cx shows additional and/or different features than
SN 1991bg, we have also included in the identifications the forbidden lines of
Fe and Ni listed in the Atomic Line List v2.04. Since our criterion for
identifying these additional lines is based purely on wavelength coincidences,
the identifications in Figure 11 (especially those marked with a ``?") are only
suggestive and should be used with caution. The true line identifications
require detailed NLTE spectral synthesis and are beyond the scope of this
paper.

Nevertheless, most of the identifications in Figure 10 are adopted from Mazzali
et al. (1997), and show that the nebular spectrum of SN 2002cx is dominated by
forbidden lines of Fe and Co. The other lines in the spectrum are the Ca~II IR
triplet, O~I $\lambda$7773, [Ca~II] $\lambda\lambda$7292, 7324, and possibly
Ca~I] $\lambda$7504. The [Ca~II] $\lambda\lambda$7292, 7324 doublet was used by
Mazzali et al. (1997) to explain the strong emission around 7300~\AA\ in SN
1991bg (perhaps with contributions from [Fe~II] $\lambda\lambda$7155, 7172 and
[Ni~II] $\lambda\lambda$7380, 7410), and the weakness of this line in SN 2002cx
is consistent with the weakness of the other Ca lines, such as the Ca~II IR
triplet and the Ca~II H \& K lines.

Since at late times the line-forming region is deep within the ejecta, the
dramatic difference between the nebular spectrum of SN 2002cx and those of
other SNe~Ia indicates that SN 2002cx has quite different composition and/or
physical conditions in the interior ejecta (temperature, density, ionization,
etc.) than other SNe~Ia. As can be seen in Figure 10, even the subluminous SN
1991bg has nebular features that match the normal SN~Ia 1994D quite well
(though narrower). The peculiar nebular spectrum of SN 2002cx may thus suggest
that it comes from a different explosion mechanism than in other SNe~Ia, though
it could also result from the extreme of one family of explosion models. The
fact that SN 1999ac shows some similarities to SN 2002cx is suggestive of the
latter hypothesis.

We also note that the spectral range 6500--8500~\AA\ corresponds to the broad
$R$ and $I$ bands, so the peculiar spectral evolution in this region may be
related to the unique photometric behavior of the SN in the $R$ and $I$ bands. In
particular, the additional emission near 7000~\AA\, may contribute to the
broad peak around maximum and the slow late-time decline in the $R$ band, and
the lack of strong Ca~II IR triplet absorption may contribute to the plateau
around maximum and the slow late-time decline in the $I$ band. More discussion of
the $R$-band and $I$-band light curves can be found in \S 4.3.

\subsection{Expansion Velocities}

The expansion velocities ($V_{exp}$), as inferred from observed minima of
absorption lines in the spectra, may provide some clue to the nature of SN Ia
explosions (Branch, Drucker, \& Jeffery 1988; Khokhlov, M\"uller, \& H\"oflich
1993). Figure 12 shows the expansion velocities derived from several lines. The
SN 2002cx and SN 1999ac data are shown with solid circles and open circles,
respectively.

SN 2002cx has the lowest expansion velocities ever reported for a SN~Ia.  For
Fe~III $\lambda$4404, Fe~III $\lambda$5129, Fe~II $\lambda$4555, and the Ca~II
H \& K lines, the measured $V_{exp}$ of SN 2002cx is $\sim$ $-$6,000 km
s$^{-1}$, only about half that of the other SNe~Ia. The $V_{exp}$ evolution is
also very flat, implying that the velocity gradient in the ejecta of SN 2002cx
is small during the period of spectral observations ($t = -$4 to 56~d).

No apparent Si~II $\lambda$6355 or S~II $\lambda\lambda$5612, 5654 lines are
observed in our spectra of SN 2002cx, but the expansion velocity measured from
these lines for other SNe~Ia are shown in Figure 12 for completeness.  Note
that SN 1999ac has the lowest $V_{exp}$ measured from S~II
$\lambda\lambda$5612, 5654 lines among the several SNe~Ia shown in Figure
12. The values of $V_{exp}$ of SN 1999ac measured from the Fe~II and Fe~III
lines are also lower than in SNe 1991T and 2000cx, though higher than in SN
2002cx.  The similarity of the $V_{exp}$ evolution between SN 2002cx and SN
1999ac is consistent with their similar photometric behavior in the early
$B$-band and $V$-band light curves, and suggests that the $V_{exp}$ evolution
and the photometric behavior of SNe~Ia may be related.

We also see that although SN 1999ac and SN 1991bg have low $V_{exp}$ measured
from S~II and Si~II lines, they have relatively normal and high $V_{exp}$
measured from the Ca~II H \& K lines.  From studies of the spectra of SN 1994D,
Hatano et al. (1999b) suggest evidence for the presence of two-component
Ca~II features, forming in high-velocity ($\gtrsim$20,000 km s$^{-1}$) and lower
velocity ($\lesssim$16,000 km s$^{-1}$) material. They further postulated that the
high-velocity gas might be ``primordial" (that is, chemically enriched prior to
the explosion), while the lower velocity ejecta might be freshly synthesized in
the SN explosion.  The lack of high-velocity calcium in SN 2002cx might thus
suggest a lack of primordial calcium, although we caution that both
components found in SN 1994D are at much higher $V_{exp}$ than in SN 2002cx, so
the two SNe may have quite different mechanisms for Ca~II line
formation. Expansion velocities of SNe~Ia are further discussed in \S 4.5.

In summary, SN 2002cx exhibits unique spectral evolution. The premaximum
spectrum is similar to that of SN 1991T and is dominated by high-excitation
Fe~III lines. The spectrum around maximum evolves very quickly and is dominated
by Fe~II lines by 2 weeks after $B$ maximum. Features from intermediate-mass
elements are weak at all times.  Mysterious emission lines are seen in
the spectral range 6500--8500~\AA, whose presence may also affect the
photometric behavior of SN 2002cx. The expansion velocities measured from the
absorption lines are also extremely low. In the next section, we discuss
how the spectral and photometric evolution of SN 2002cx are related, and we
explore possible models for this peculiar object.

\section{Discussion}

\subsection{Is SN 2002cx a Type Ia Supernova?}

  Given the strange observed properties of SN 2002cx, it is legitimate to
question whether the SN~Ia classification for SN 2002cx is secure.  According
to the commonly accepted classification criteria of SNe~Ia [that is,
hydrogen-deficient SNe whose near-maximum spectra show conspicuous absorption
features near 6150~\AA\ (due to Si~II) and near 3750~\AA\ (due to Ca~II), as
well as absorption features from S~II, O~I, Fe~II, and Co~II], it is uncertain
that SN 2002cx is a SN~Ia; the Si~II (especially Si~II $\lambda$6355) and S~II
features have not been successfully observed, in part due to our
sparse observations and the extremely fast spectral evolution around maximum
brightness. 

   However, we consider the SN~Ia classification for SN 2002cx to be secure,
because its photometric and spectral evolution can be grossly explained within
the paradigm of SN~Ia observations (\S 2 and \S 3), though with considerable
differences.  The premaximum spectra resemble those of the peculiar SN~Ia
1991T. The spectrum at $t$ = 12~d is similar to those of other SNe Ia, though
the Fe~II lines are more prominent. In addition, the late-time spectra are
dominated by emission lines of iron and cobalt, reminiscent of SNe~Ia but not
of other SN types. Finally, the data suggest that SNe 2002cx and 2000cx may be
extreme ends of the same class of objects, and SN 2000cx is clearly a SN~Ia.

\subsection{Absolute Magnitudes}

The host galaxy of SN 2002cx, CGCG 044-035, has a heliocentric radial velocity
of 7184 km s$^{-1}$ as reported in NED (and also confirmed by the narrow
emission lines seen in our spectra from H~II regions in the host galaxy), so it
lies in the Hubble flow and we can use its velocity to derive a reliable
distance. Converting $v = cz$ to a velocity in the cosmic microwave background
(CMB) frame according to the prescription of de Vaucouleurs et al. (1991), we
find $v_{CMB} = 7489 $ km s$^{-1}$.

  Adopting the final result from the {\it Hubble Space Telescope (HST)} $H_0$
Key Project (Freedman et al. 2001), $H_0 = 72 \pm 8$ km s$^{-1}$ Mpc$^{-1}$,
the distance modulus of SN 2002cx is $\mu = (m - M) = 35.09\pm 0.32$ mag from
the $v_{CMB}$ derived above; the uncertainty is a combination of the error bar
in $H_0$ and a possible peculiar motion of 700 km s$^{-1}$ for the host
galaxy\footnote{The host galaxy of SN 2002cx, CGCG 044-035, is in a poor
cluster WBL 435 (White et al. 1999) which consists of 6 members. The mean
radial velocity of the cluster (with only 3 members have measurements) is
$v_{CMB} = 6810 $ km s$^{-1}$. The rather high peculiar velocity we adopt for
CGCG 044-035 (700 km s$^{-1}$) is about the same as the difference between the
$v_{CMB}$ measurements of the cluster and CGCG 044-035.}.  Using the apparent
peak magnitudes as listed in Table 3 and our estimate of the reddening, we
derive the peak absolute magnitudes for SN 2002cx in all filters (Table 5). The
quoted uncertainties are the sums in quadrature of the uncertainties in peak
magnitude and distance. For comparison, we also list the absolute magnitudes of
SNe 1991T, 2000cx, 1999ac, 1994D, 1991bg, and a ``standard" SN~Ia with $\Delta
m_{15} = 1.1$ mag (Gibson \& Stetson 2001).

It can be seen from Table 5 that SN 2002cx is subluminous: it is $\sim 1.8$ mag
fainter than the standard SN~Ia in the $B$ and $V$ bands. Compared to SN
1991bg, the least luminous SN~Ia known to date, SN 2002cx is brighter by $\sim
1$ mag in the $B$ band, by $\sim 0.3$ mag in the $V$ band, and is comparable in
brightness in the $R$ and $I$ bands.  SN 2002cx, SN 1994D, and SN 1999ac have
similar $\Delta m_{15}(B)$ measurements (\S 2), but SN 2002cx is less luminous
than the other two SNe by 1.4 mag in the $B$ band, and is also fainter in the
other bands.

We explored alternative causes for the faint apparent absolute magnitudes of SN
2002cx (e.g., SN 2002cx occurs in a background galaxy, high reddening for SN
2002cx, huge peculiar radial velocity), but we found that they conflict with
the observations in one way or another. We also postulate that SN 2002cx is
unlikely to have a normal luminosity, given its peculiar photometric behavior
in the $R$ and $I$ bands and its unique spectral evolution. We thus conclude
that SN 2002cx is intrinsically subluminous.

\subsection{The $R$-Band and $I$-Band Secondary Maxima}

SN 2002cx exhibits peculiar evolution around the $R$-band and $I$-band
secondary maxima: there is a broad peak in the $R$ band, and a plateau in the
$I$ band that lasts until the phase at which the secondary maximum of other
SNe~Ia occurs. In this section we explore possible reasons for this peculiar
behavior.

The IR secondary maximum in SNe~Ia is explained by H\"oflich, Khokhlov, \&
Wheeler (1995; hereafter HKW95) as an effect produced by the effective
temperature ($T_{eff}$) and photospheric radius ($R_{ph}$) --- if $R_{ph}$ is
still increasing well after maximum light and $T_{eff}$ is decreasing slowly, a
secondary maximum is formed.  A long-term increase of $R_{ph}$ requires high
opacity, and hence high temperature in the outer region since the opacity in
SNe~Ia drops drastically when the temperature falls below 20,000~K (H\"oflich,
M\"uller, \& Khokhlov 1993).  The lack of secondary IR maxima in the
subluminous SNe~Ia (SN 1991bg-like objects) is explained in terms of a receding
$R_{ph}$ soon after maximum brightness, which leads to a merging of the
``first" and ``second" maxima.  (However, the models shown by HKW95 only have a
secondary maximum in the $J$ and redder bands.) In SN 2002cx, the expanding
fireball effect is not an attractive explanation for the $I$-band plateau
phase: the spectrum evolves rapidly during the first 2 weeks after $B$ maximum,
probably inconsistent with a nearly constant temperature in the ejecta, and
$R_{ph}$ is likely to be decreasing because the spectrum rapidly approaches the
nebular phase.

   Suntzeff (1995) argues that the $RI$ secondary maximum in SNe~Ia is likely
to be caused by an overall redistribution of continuum flux from blue to red,
rather than by a few isolated spectral features. This is similar to the
explanation of Pinto \& Eastman (2000), who suggest that the secondary maximum
is an ionization/opacity effect. When the photosphere recedes, the ionization
in regions of trapped radiation falls and includes significant amounts of
singly ionized species (Ca~II, Fe~II, Co~II) that can emit strongly in the
near-IR.  This leads to a sharp reduction in the flux mean opacity.  The
diffusion time is thus reduced, and the residual stored energy is released,
leading to the IR secondary peak.  (However, these authors do not discuss in
detail why this mechanism works for normal and overluminous SNe~Ia, but not for
subluminous objects.) The ionization/opacity effect is an attractive explanation
for the $RI$-band evolution of SN 2002cx, and for the comparison with SN
2000cx. The rapid change in the ionization stage (from Fe~III to Fe~II) in SN
2002cx may result in an early release of the residual stored energy, and thus a
merging of the secondary peak with the first one. Consequently, a broad peak in
the $R$ band and a plateau in the $I$ band may occur. In contrast, SN 2000cx
has a slow change in the ionization stage (from Fe~III to Fe~II), and it
exhibits a prominent ``dip" and strong secondary maximum in the $I$ band (Li et
al. 2001b).

The peculiar spectral evolution in the range 6500--8500~\AA\ suggests that at
least for SN 2002cx, spectral features play some role in the formation of the
$RI$ secondary maximum. In particular, the additional features around 7000~\AA\
may contribute to the broad peak in the $R$ band, and the lack of a strong
Ca~II IR triplet may contribute to the plateau phase in the $I$ band.

In summary, the ionization/opacity effect and the spectral feature effect may
contribute to the $R$-band and $I$-band evolution around the secondary maximum
in SN 2002cx; the fireball effect is not expected to be an important
factor. Detailed modeling of both the light-curve shape and the spectral
evolution, beyond the scope of this paper, is required to study the relative
contributions of different effects.

\subsection{The Late-Time Decline Rate}

There is evidence that SN 2002cx has a small ejecta mass and small $^{56}$Ni
production: the rapid initial decline after maximum in the $B$ band, the fast
spectral evolution after maximum brightness, the early transition to the
nebular phase, and the low luminosity. It has been shown from both light curve
and spectral modeling (HKW95; Mazzali et al. 2001) that the luminosity of a
SN~Ia is tightly correlated with the $^{56}$Ni produced in the explosion. Thus,
SN 2002cx should have synthesized an amount of $^{56}$Ni similar to that of SN
1991bg ($\sim 0.1~M_\odot$; Mazzali et al. 1997) because they have similar
absolute magnitudes (though SN 2002cx is slightly more luminous).

Assuming the galaxy background does not significanly contaminate the late-time
light curves, SN 2002cx has very slow late-time decline rates in all $BVRI$
passbands\footnote{The slow late-time decline is also confirmed by the recovery
of SN 2002cx in images taken at $t \approx 7$ months. We estimate that SN
2002cx has $I < 20.5$ mag at $t$ = 214~d, which suggests that it declined at a
rate slower than 0.015 mag d$^{-1}$ in the $I$ band. This will be investigated
further in the future, when template images devoid of significant SN light are
subtracted from the earlier data}. At first glance, this seems inconsistent
with the small ejecta mass and small $^{56}$Ni production inferred above:
smaller ejecta mass leads to a shorter diffusion time and a less efficient
trapping of the $\gamma$-ray photons from the radioactive decay of $^{56}$Ni,
and less $^{56}$Ni yields a lower temperature and thus reduced mean opacity.
Both of these trends produce an atmosphere that cools more rapidly, a
photosphere that recedes faster in mass coordinates, and a luminosity that
declines faster. However, as suggested by Pinto \& Eastman (2002) and HKW95, a
very slow expansion velocity (as seen in SN 2002cx) has a countereffect: it
results in a smaller escape probability for $\gamma$-ray photons and longer
diffusion timescales, and hence slow late-time decline.

The fact that both SN 2002cx and SN 1991bg have small ejecta masses, but the
former has slow late-time decline while the latter has fast late-time decline,
suggests that the late-time photometric evolution of SN 2002cx may be dominated
by the $V_{exp}$ effect, while that of SN 1991bg may be dominated by the ejecta
mass effect. Note that SN 2002cx shows a trend that is consistent with SN
2000cx, which has a high $V_{exp}$ and fast late-time decline.

An alternative explanation for the slow late-time decline is the positron
transport. As discussed by Milne, The, \& Leising (2001), the late-time energy
deposition in a SN~Ia is dominated by positron transport, which depends on the
nature of the magnetic field. One scenario invokes a strong magnetic field that
is turbulently disordered such that positrons mirror frequently with no net
transport (the trapping scenario; Axelrod 1980).  Although Milne et
al. suggested that existing late-time observations of a sample of 22 SNe~Ia are
more consistent with models having substantial positron escape, SN 2002cx may
be peculiar and is more consistent with the trapping scenario. We note that for
most SNe~Ia, however, different positron transport scenarios show significant
differences only in late-time decline starting about 300~d after the explosion,
considerably later than our observations of SN 2002cx.

\subsection{Expansion Velocities}

We first note that expansion velocities ($V_{exp}$) inferred from the observed
Doppler-shifted absorption lines may not exactly follow the photospheric
velocities ($V_{ph}$), as has been shown in published theoretical models, since
the former depend on the abundances as a function of depth, while the latter
are affected by the expansion velocity of the ejecta, the density structure,
and the composition structure (e.g., the distribution of $^{56}$Ni). Since
lines are usually formed somewhere above the photosphere, $V_{exp}$ is
generally an overestimate of $V_{ph}$.

There is no reason why $V_{ph}$ and $V_{exp}$ should correlate with $^{56}$Ni
production and thus with luminosity of SNe~Ia (Hatano et al. 2000), since the
velocities also depend on other factors such as the density and temperature
structure in the ejecta. For instance, SN 2000cx has high $V_{exp}$ but a
relatively normal luminosity, while SN 1991T has a normal $V_{exp}$ but is
relatively overluminous (Figure 12 and Table 5). However, some combinations of
$V_{exp}$ and luminosity seem to be forbidden, at least in the current sample
of observed SNe~Ia: neither a SN~Ia with high $V_{exp}$ and low luminosity nor
a SN~Ia with low $V_{exp}$ and high luminosity has been observed.

SN 2002cx has the lowest $V_{exp}$ ever measured for a SN~Ia, only $\sim$50\%
of that of other SNe~Ia. The evolution of $V_{exp}$ is also rather flat (\S
3). Khokhlov et al. (1993) suggest that the $V_{ph}$
evolution is model dependent (see \S 4.7 for more discussion of SN~Ia
models). For example, in detonation and delayed detonation models $V_{ph}$
decreases monotonically, in pulsating delayed detonation and envelope models
$V_{ph}$ has a plateau phase after maximum, while in deflagration models
$V_{ph}$ has a plateau before maximum and a monotonically decreasing phase
after maximum. Although the $V_{exp}$ evolution of SN 2002cx is mostly
consistent with the pulsating delayed detonation and envelope models, we
emphasize that $V_{exp}$ is different from $V_{ph}$ as discussed above, and
that the suggestion of Khokhlov et al. (1993) is based on much higher $V_{ph}$
than those measured in SN 2002cx, so the $V_{exp}$ evolution alone will not
single out any model for SN 2002cx. On the other hand, a successful theoretical
model for SN 2002cx will need to address the extremely low expansion velocity
and its flat evolution.

The fact that both a high $V_{exp}$ event such as SN 2000cx, and a low
$V_{exp}$ event such as SN 2002cx, show peculiar photometric behavior suggests
that $V_{exp}$ may be a second parameter that demonstrates the diversity of
SNe~Ia\footnote{The color of SNe~Ia at maximum has been used as a second
parameter to describe the SN~Ia family (e.g., Tripp \& Branch 1999; Drenkhahn
\& Richtler 1999), but these methods suffer from the currently poor
understanding of the intrinsic colors of SNe~Ia and the difficulties of
measuring the host-galaxy reddening.}. Although the absolute magnitudes and
$V_{exp}$ do not seem to be correlated (e.g., Wells et al. 1994; Patat et
al. 1996), $V_{exp}$ may indicate other properties of the light curves. For
instance, the initial decline rate [$\Delta m_{15}(B), ~\Delta m_{15}(V)$] might
need to be modified according to the $V_{exp}$ of each SN, so that the decline
rate vs. luminosity correlation has a smaller dispersion and can also be
applied to events with unusually high or low $V_{exp}$. The details of the
relation between $V_{exp}$ and photometric behavior of SNe~Ia will be examined
in a future paper.

\subsection{SN 2002cx and the Observational Correlations}

SN 2002cx is a counterexample to the connection between spectral and
photometric behavior of SNe~Ia. For example, (1) SNe~Ia having SN 1991T-like
premaximum spectra are expected to be overluminous, but SN 2002cx has SN
1991T-like spectra and is subluminous; (2) subluminous SNe~Ia such as SN 1991bg
usually have strong Si~II and Ca~II lines, but in SN 2002cx these lines are
very weak.

SN 2002cx may also be inconsistent with the important luminosity vs. decline
rate relation, which is the cornerstone of several empirical luminosity
correction techniques such as the $\Delta m_{15}(B)$ method (Phillips 1993;
Hamuy et al. 1996a; Phillips et al. 1999), the multicolor light curve shape
method (Riess, Press, \& Kirshner 1996; Riess et al. 1998; Jha 2002), and the
stretch method (Perlmutter et al. 1997). To some extent this does, however,
hinge on the result of $\Delta m_{15}(B)$ from the future final photometry
using template galaxy subtractions.  Assuming the final $\Delta m_{15}(B)$ is
not significantly changed, and using $\Delta m_{15}(B) = 1.29\pm0.11$ mag for
SN 2002cx as derived in \S 2, the prescription of Phillips et al. (1999) for
the $\Delta m_{15}(B)$ method, and the absolute magnitudes for a standard SN~Ia
as listed in Table 5, the expected apparent absolute magnitude for SN 2002cx is
$M_B = 19.15\pm0.10$ mag, and $M_V = 19.16\pm0.08$ mag, much brighter (by $\sim
1.6$ mag) than the actual values of SN 2002cx in Table 5.  This inconsistency
is also qualitatively shown in \S 4.2, where it is demonstrated that SN 2002cx,
SN 1994D, and SN 1999ac have similar $\Delta m_{15}(B)$ values, but SN 2002cx
is fainter by about 1.4 mag in the $B$ band than the other two SNe~Ia.

So far no other SNe~Ia are known with properties similar to those of SN 2002cx,
so it may be unique, just as SN 2000cx is unique, but in an opposite way.
Fortunately, these two counterexamples to the observed correlations are
distinguished by their peculiar expansion velocities, so one is advised to
check both the spectral peculiarities and the expansion velocity of a SN~Ia
before applying the correlations.

\subsection{Theoretical Models for SN 2002cx}

The observations of SN 2002cx offer the following constraints on its possible
progenitors and explosion scenarios. (1) Peculiar photometric evolution,
especially in the $R$ and $I$ bands. (2) Peculiar color evolution: blue 
$(B - V)_0$ color at maximum, red $(V - R)_0$ and $(V - I)_0$ evolution. (3)
Low luminosity at all optical wavelengths. (4) Iron-group elements in
the hot outer ejecta. (5) Little production of intermediate-mass elements
in the explosion. (6) Small ejecta mass, and rapid transition to the
nebular phase. (7) Low expansion velocity and kinetic energy.

Armed with these constraints, we begin our journey in search of a plausible
theoretical model for SN 2002cx. There is general agreement that SNe~Ia result
from thermonuclear explosions of degenerate white dwarfs (WDs; Hoyle \& Fowler
1960). Within this broad picture, three explosion scenarios dominate the
present considerations.  (1) A C-O WD near the Chandrasekhar mass accretes H or
He from a binary companion until it reaches a mass at which the central carbon
ignites ($M_{Ch}$ WD scenario; Whelan \& Iben 1973). (2) Two low-mass WDs in a
binary system merge: the smaller, more massive WD accretes material from a disk
formed by the total disruption of the less-massive companion and explodes
(merging scenario; Webbink 1984; Iben \& Tutukov 1984). (3) A low-mass
(sub-$M_{Ch}$) C-O WD accretes a He shell, which becomes thick enough to
produce a He shell detonation; this, in turn, triggers central carbon ignition
(He detonation scenario; Nomoto 1980; Woosley, Weaver, \& Taam 1980).  Within
the $M_{Ch}$ WD and the merging scenarios, the description of the propagation
of the burning front further divides the models into several groups such as
deflagration (Nomoto, Thielemann, \& Yokoi 1984), detonation (Arnett 1969),
delayed detonation (DD; Khokhlov 1991a), pulsating delayed detonation (PDD;
Khokhlov 1991b), and late detonation (Yamaoka et al. 1992).

The merging scenarios were studied by H\"oflich et al. (1993)
and found to be generally consistent with events having slow light-curve
evolution such as SN 1991T; thus, they are not viable models for SN
2002cx. However, the variety of these models has not been fully explored, and
the final result of a merging configuration (i.e., whether it explodes to
produce a SN~Ia, or collapses to form a neutron star) is still being debated
(e.g., Mochkovitch \& Livio 1990; Saio \& Nomoto 1998).

The $M_{Ch}$ WD scenario is currently the most favored model for SNe~Ia (Branch
2001). In particular, the DD and PDD models have been found to reproduce the
optical and IR light curves and spectra of SNe~Ia reasonably well (H\"oflich
1995; HKW95; Nugent et al. 1997; H\"oflich, Wheeler, \& Thielemann 1998). In
these models, the variation in the parameter ($\rho_{tr}$) at which the
transition is made from deflagration to detonation gives a range of $^{56}$Ni
mass. Models with smaller $\rho_{tr}$ give less nickel and hence both lower
peak luminosity and lower temperatures. The latter gives lower opacity and
hence a steeper decline in the light curve.  Within this hypothesis, typical
features of subluminous models are the greater production of O, S, and Si
compared to SNe~Ia of normal luminosity, and $^{56}$Ni is constrained to the
inner layers with low expansion velocities which become visible a few weeks
after maximum light. H\"oflich et al. (2002) also indicated that strong O~I
lines are expected in subluminous SNe~Ia as a result of massive oxygen layers
caused by carbon burning. Although some features of these subluminous models
contradict the SN 2002cx observations (e.g., more IME production, centrally
distributed $^{56}$Ni, and no particularly strong O~I lines have been observed
in SN 2002cx), our attention was drawn to one of these models, PDD535 (HKW95).

The model PDD535 has a rather small $\rho_{tr}$, and it produces only 0.15
M$_\odot$ of $^{56}$Ni. The final kinetic energy is only a quarter that of most
other models. The average expansion velocity in the ejecta is $\sim$ 4,500 km
s$^{-1}$. This is the only model we found in the literature that has low
$V_{exp}$ comparable to that observed in SN 2002cx.  This model is also
subluminous ($M_V = -17.77$ mag), with a risetime of 21.3~d in the $V$ band and
a $(B - V)$ color at maximum of 0.6 mag. Thus, the model successfully explains
the low luminosity, low $V_{exp}$, low kinetic energy, and slow late-time
decline rates in SN 2002cx. However, its IME production is too large, it does
not have Fe-group elements in the outer ejecta, and its $(B - V)$ color is too
red at maximum brightness.

The He detonation scenarios have been investigated by a number of
studies. Woosley \& Weaver (1994; hereafter WW94) showed that these models
synthesize less $^{56}$Ni, and have lower kinetic energy, than the $M_{Ch}$ WD
models.  Their models are also subluminous (fainter by $\sim 1$ mag than a
normal SN~Ia), though HK96 suggested that He detonations could be as bright as
other explosion scenarios. The models studied in WW94 also predict
overproduction of $^{44}$Ti, which probably contributes to the formation of
the Ti~II absorption line
around 4100--4400~\AA\ observed in the subluminous SN 1991bg\footnote{However,
Mazzali et al. (1997) showed that the Ti~II lines are likely caused by the low
temperature in the ejecta rather than a high abundance of $^{44}$Ti, since even
with the W7 deflagration abundance the synthetic spectrum for SN 1991bg shows
strong Ti~II lines.}. The $^{56}$Ni is distributed in the outer shell, and
smaller quantities of IME are produced in the explosion than in the $M_{Ch}$
WD models.

The He detonations are not considered to be good general models for SNe~Ia, as
they show either a weak (WW94) or opposite (HK96) decline rate vs. luminosity
relation. The decline rate is too fast for normal SNe~Ia, and the color at
maximum is too blue for subluminous SNe~Ia such as SN 1991bg (but perhaps not
for SN 2002cx, at least in $B-V$).  Moreover, both one-dimensional and
two-dimensional calculations for these models predict that the Si-rich layers
are restricted to velocities smaller than 14,000 km s$^{-1}$ (WW94; Livne \&
Arnett 1995; HK96), while observations indicate expansion velocities of Si-rich
layers in excess of 20,000 km s$^{-1}$ (e.g., SN 1990N, Leibundgut et al. 1991;
SN 1994D, H\"oflich 1995).

Nugent et al. (1997) investigated the synthetic spectra of the He detonation
models, and suggested that although they bear some resemblance to the
subluminous SNe~Ia such as SN 1991bg, they don't have the observed strong UV
deficit and they have extremely weak IME features. Helium lines are not
prominent in the synthetic spectra near maximum, despite the significant He
abundance.

The He detonations are in some respects attractive models for SN 2002cx. The
general difficulties of the models to explain subluminous SNe~Ia as discussed
above (e.g., too blue at maximum, weak IME features, and Fe-group elements in
the outer ejecta) do not apply to the observations of SN 2002cx. The most
promising model for SN 2002cx that we found in this category is the HeD6 model
discussed by HK96, which consists of a 0.6~$M_\odot$ C-O WD and a
0.17~$M_\odot$ He layer. The final kinetic energy is about half that of most
other models and 0.25~$M_\odot$ of $^{56}$Ni is synthesized in the explosion.
The model has $M_V = -18.56$, $(B - V) = 0.05$ mag at maximum, a risetime of
14.7 days in the $V$ band, and $\Delta m_{15}(V) = 1.5$ mag.  The successful
aspects of this model for explaining SN 2002cx are as follows: (1) Fe-group
elements are distributed in the outer part of the ejecta, (2) blue $(B - V)$
color at maximum, and (3) no strong IME features in the spectra. On the other
hand, detracting from its success, (1) the model still predicts high expansion
velocity for Fe-group elements, and generally higher $V_{exp}$ for other
elements than those observed in SN 2002cx, (2) the model light curves decline
too fast, (3) there are secondary maxima in the $R$ and $I$ bands, and (4) the
model is still too luminous.

Earlier studies of the He detonations (Woosley et al. 1980; Nomoto
1982) suggested the possibility of a faint SN in which only the He layer burned
and a white dwarf remained after the explosion (i.e., a ``super nova" instead
of a real supernova). This was subsequently criticized by Livne (1990) as
overlooking the possibility that the ingoing compressional wave might ignite an
outwardly propagating carbon detonation at or near the center, which is
confirmed by two-dimensional studies by Livne \& Glasner (1991). On the other
hand, WW94 emphasize that three-dimensional simulations are necessary to study
the final outcome of the He detonations. While SN 2002cx may be a peculiar
event with only the He layer burned and a white dwarf remaining, light curves
and spectral syntheses of this kind of explosion are not yet available for a
detailed comparison with our observations.

Nomoto et al. (1995) discussed accretion-induced collapse (AIC) of WDs and
suggested that the subluminous SN 1991bg might be produced by AIC after the
merging of two WDs. The idea is that the merging of an O-Ne-Mg WD with a C-O WD
results in the collapse of the O-Ne-Mg WD and the explosion of the outer C-O
envelope with the ejection of $\sim 0.6 M_\odot$ of material.  However, the
ejecta in these models move at much higher expansion velocities (10,000 to
15,000 km s$^{-1}$) than those observed in SN 2002cx.

We conclude that not a single published theoretical model successfully explains
all observed aspects of SN 2002cx. We also note that the dramatic differences
between the two subluminous SNe~Ia 2002cx and 1991bg (the initial decline rate
in the $B$ and $V$ bands, the evolution around the $R$-band and $I$-band
secondary maximum, the late-time decline rates, the premaximum spectra, the
Fe~III, Si~II, Ca~II, and Ti~II line evolution, the nebular spectra, and the
$V_{exp}$) suggest that they arise from quite different explosion mechanisms.
Since most of the current theoretical studies have been geared toward the
explanation of objects like SN 1991bg, it is not surprising that we were unable
to find a successful model for SN 2002cx.  Additional models and explosion
scenarios may need to be constructed to explain the challenge provided by SN
2002cx.

SN 2002cx adds to the diversity already witnessed among SNe~Ia.  It seems that
the observed quantities of SNe~Ia (e.g., luminosity, initial decline rate such
as $\Delta m_{15}(B)$, secondary maximum, spectral evolution, and expansion
velocity) require the models of SNe~Ia to have a range of explosion energy,
kinetic energy, envelope mass, and composition structure. While it is possible
to vary parameters within one family of explosion models to simulate the
majority of SNe~Ia, the observations of SN 2002cx suggest that at least for the
subluminous end, it is likely that multiple channels are required. Notice that
HK96 also concluded that several explosion mechanisms are required to explain
the variations among SNe~Ia.

\section{Conclusions}

(1) SN 2002cx is a bizarre object --- indeed, unique among all known SNe~Ia. It
has a premaximum spectrum similar to that of SN 1991T, but a luminosity similar
to that of SN 1991bg. The expansion velocity measured from the spectra is the
lowest ever recorded for a SN~Ia.

(2) The photometric evolution of SN 2002cx is peculiar. In the $B$ band it has
a decline rate of $\Delta m_{15}(B) = 1.29\pm0.11$ mag, similar to those of SN
1994D and SN 1999ac, but it is fainter by $\sim 1.4$ mag than SN 1994D and SN
1999ac. The $R$ band has a very broad peak, and the $I$ band has a unique
plateau that lasts until about 20 days after $B$ maximum. The late-time decline
rate is rather slow in all $BVRI$ bands.  The $(B - V)$ color evolution is
nearly normal, but the $(V - R)$ and $(V - I)$ colors are very red.

(3) The premaximum spectrum of SN 2002cx resembles those of SN 1991T-like
objects, but with extremely low expansion velocities. The spectral evolution is
dominated by Fe-group element lines, with very weak intermediate-mass element
features. There are mysterious emission lines near 7000~\AA\ around 3 weeks
after maximum brightness. The nebular phase, which was reached quite soon after
maximum, is also very different from those of other SNe~Ia.

(4) SN 2002cx is inconsistent with the observed spectral vs. photometric
sequence, and also quantitatively with the SN~Ia decline rate vs. luminosity
relation.  No existing theoretical model successfully explains all observed
aspects of SN 2002cx, though the pulsating delayed detonation of a $M_{Ch}$ WD
or the He detonation of a sub-$M_{Ch}$ WD have some promising characteristics
and should be pursued further.

\acknowledgments

We thank the Lick Observatory, Keck Observatory, and FLWO staffs for their
assistance, W. M. Wood-Vasey for his early alert of the discovery of SN 2002cx,
and S. R. Kulkarni for the Keck observation of SN 2002cx on 2002 June 15. The
W. M. Keck Observatory is operated as a scientific partnership among the
California Institute of Technology, the University of California, and NASA; the
observatory was made possible by the generous financial support of the
W. M. Keck Foundation.  The work of A.V.F.'s group at U. C. Berkeley is
supported by National Science Foundation grant AST-9987438, as well as by the
Sylvia and Jim Katzman Foundation. Additional funding is provided by NASA
through grants GO-9114 and GO-9428 from the Space Telescope Science Institute,
which is operated by the Association of Universities for Research in Astronomy,
Inc., under NASA contract NAS~5-26555. KAIT was made possible by generous
donations from Sun Microsystems, Inc., the Hewlett-Packard Company, AutoScope
Corporation, Lick Observatory, the National Science Foundation, the University
of California, and the Katzman Foundation.

\newpage

\renewcommand{\baselinestretch}{1.0}

\newpage

\begin{figure}
{\plotfiddle{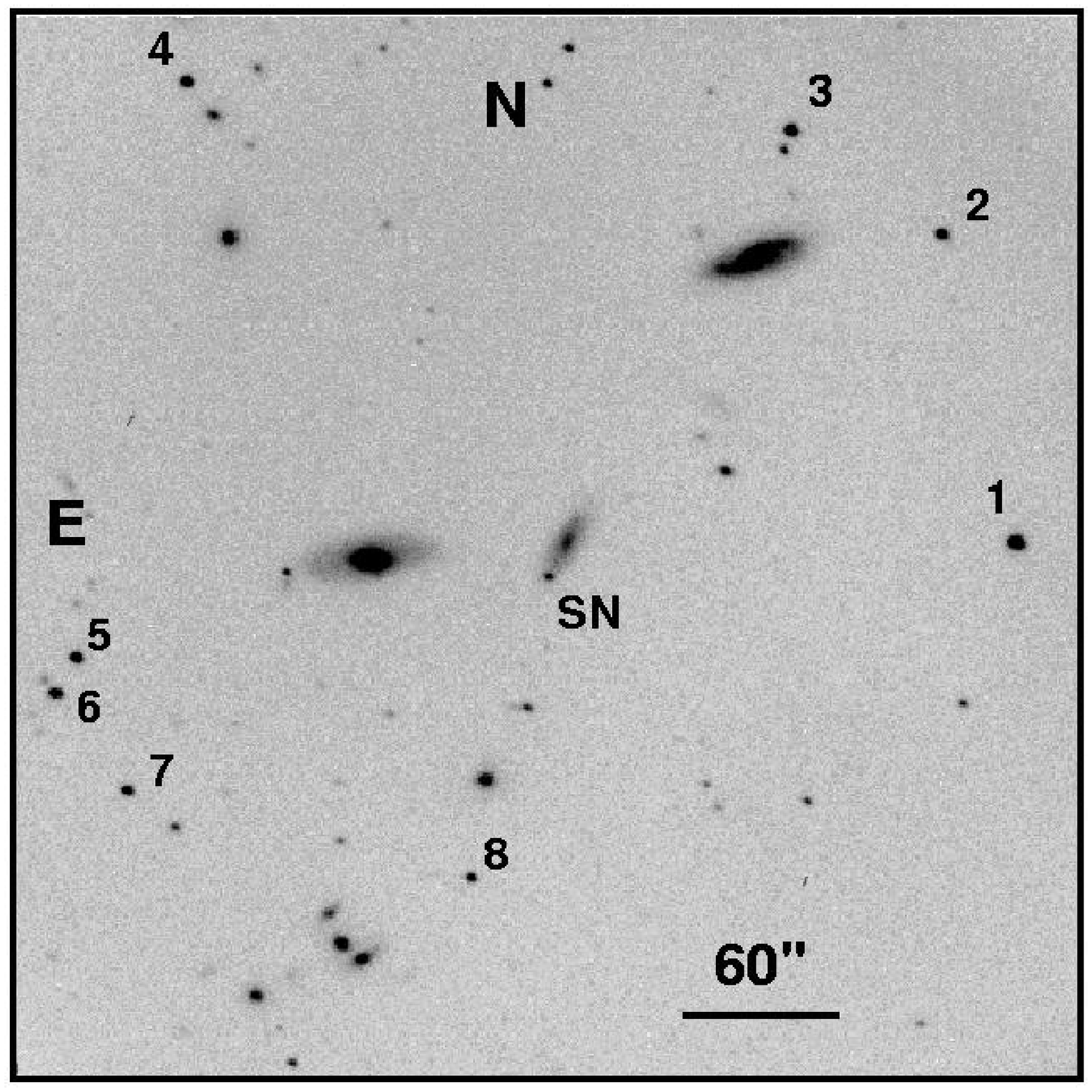}{6.4in}{0}{90}{90}{-30}{-60}}
\caption{\emph{V}-band KAIT image of the field of SN 2002cx, taken on 
2002 May 18. The field of view is 6$\farcm$7 $\times$ 6$\farcm$7. The
eight local standard stars are marked (1--8).}
\label{1}
\end{figure}

\newpage

\begin{figure}
{\plotfiddle{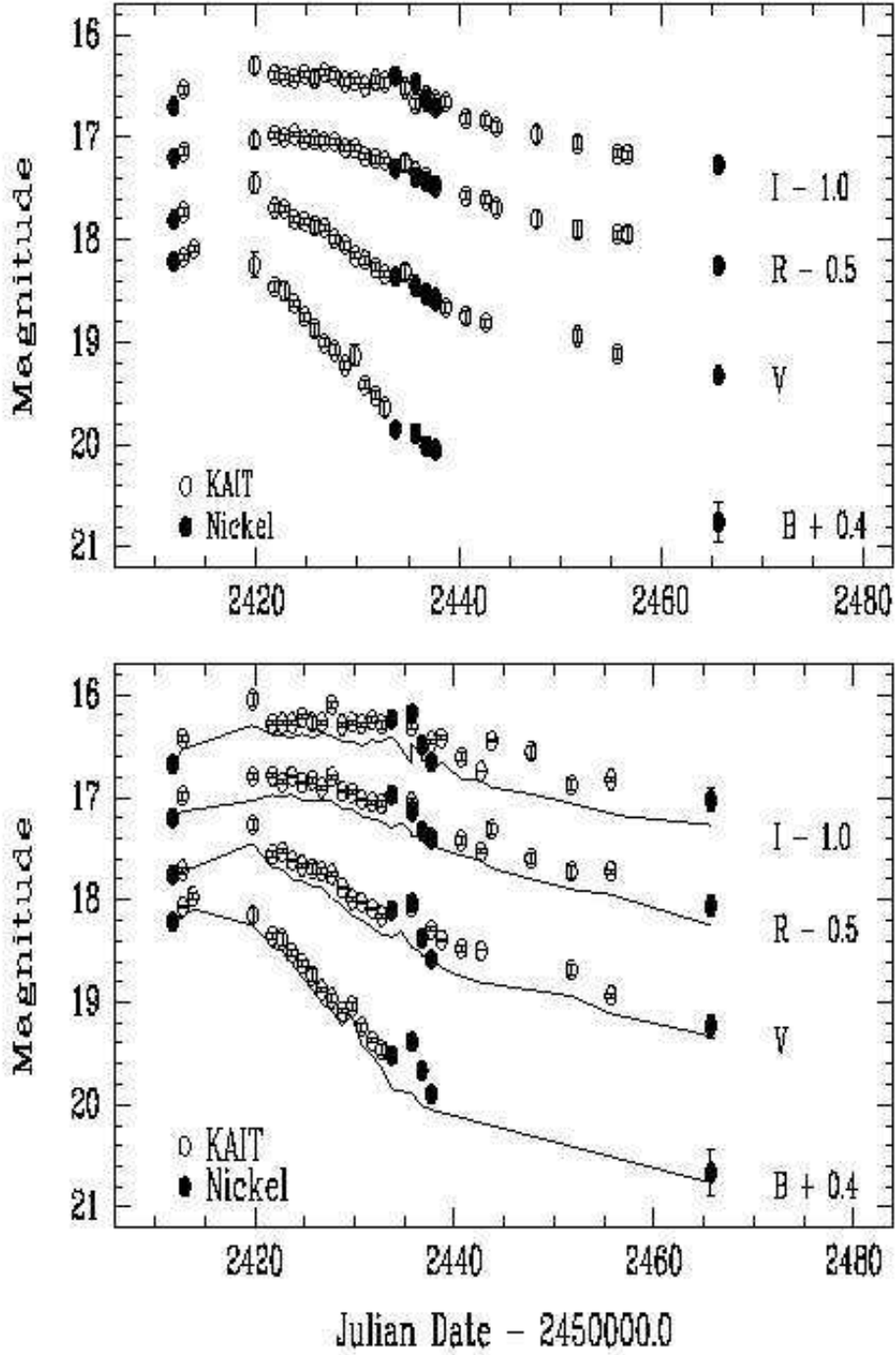}{6.4in}{0}{90}{90}{-30}{-60}}
\caption{The preliminary \emph{B, V, R,} and \emph{I} light curves of SN 2002cx. 
The open circles are the KAIT measurements and the solid circles are the 
Nickel data. For most of the points the statistical uncertainties are smaller than
the plotted symbols. The upper panel shows the results from the adopted
galaxy-subtraction technique discussed in the text, while the lower panel
shows a comparison between the galaxy-subtraction photometry (solid lines)
and the PSF-fitting photometry (open and solid circles).}
\label{2}
\end{figure}

\begin{figure}
{\plotfiddle{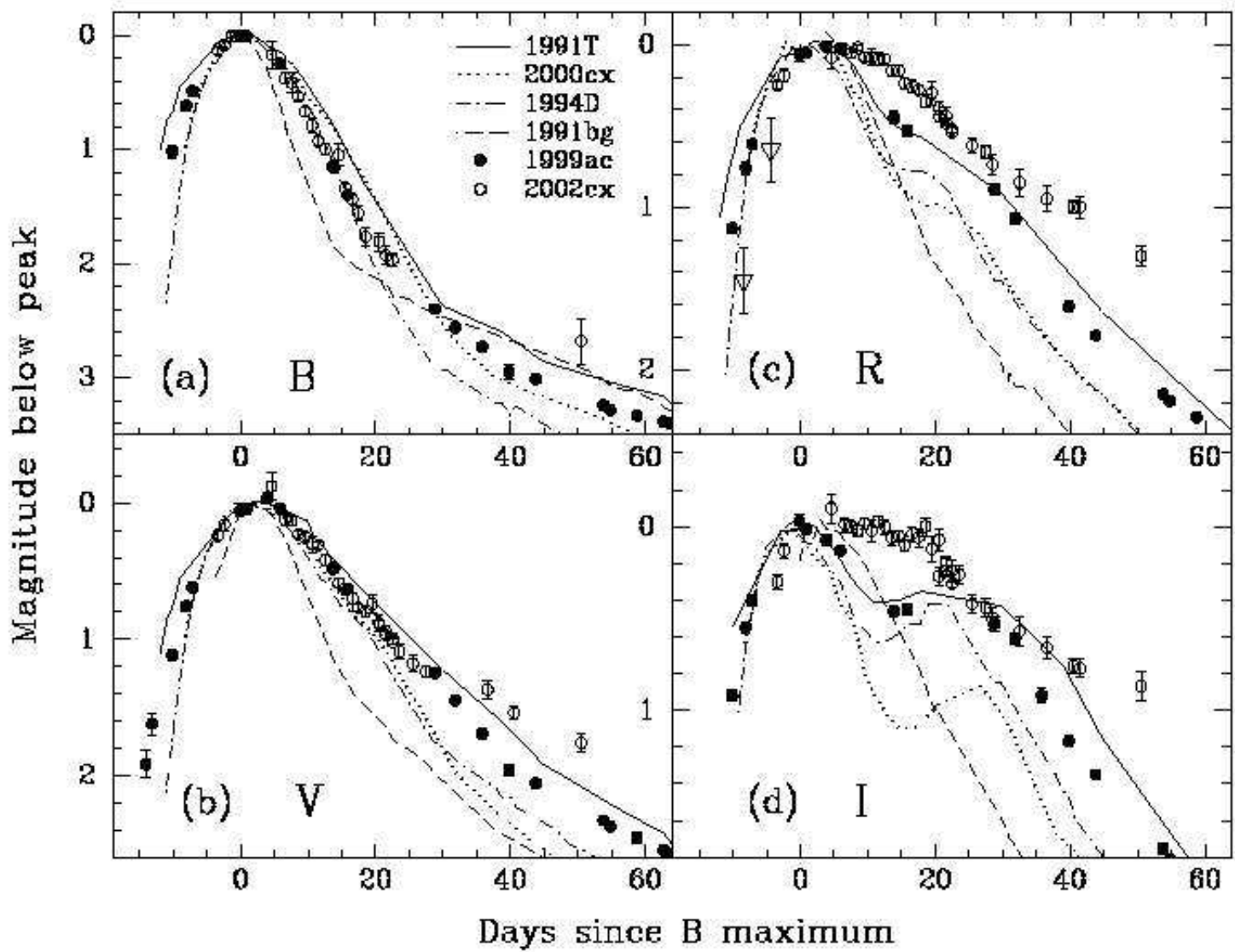}{6.4in}{-90}{90}{90}{-100}{560}}
\caption{Comparison between the preliminary \emph{B, V, R,} and \emph{I} light curves 
of SN 2002cx and those of SN 1991T (Lira et al. 1998), SN 2000cx (Li et al.
2001b), SN 1994D (Richmond et al. 1995), SN 1991bg (Filippenko et al. 1992a;
Leibundgut et al. 1993), and SN 1999ac (Li et al. in preparation). 
All light curves are shifted
in time and peak magnitude to match those of SN 2002cx.}
\label{3}
\end{figure}

\begin{figure}
{\plotfiddle{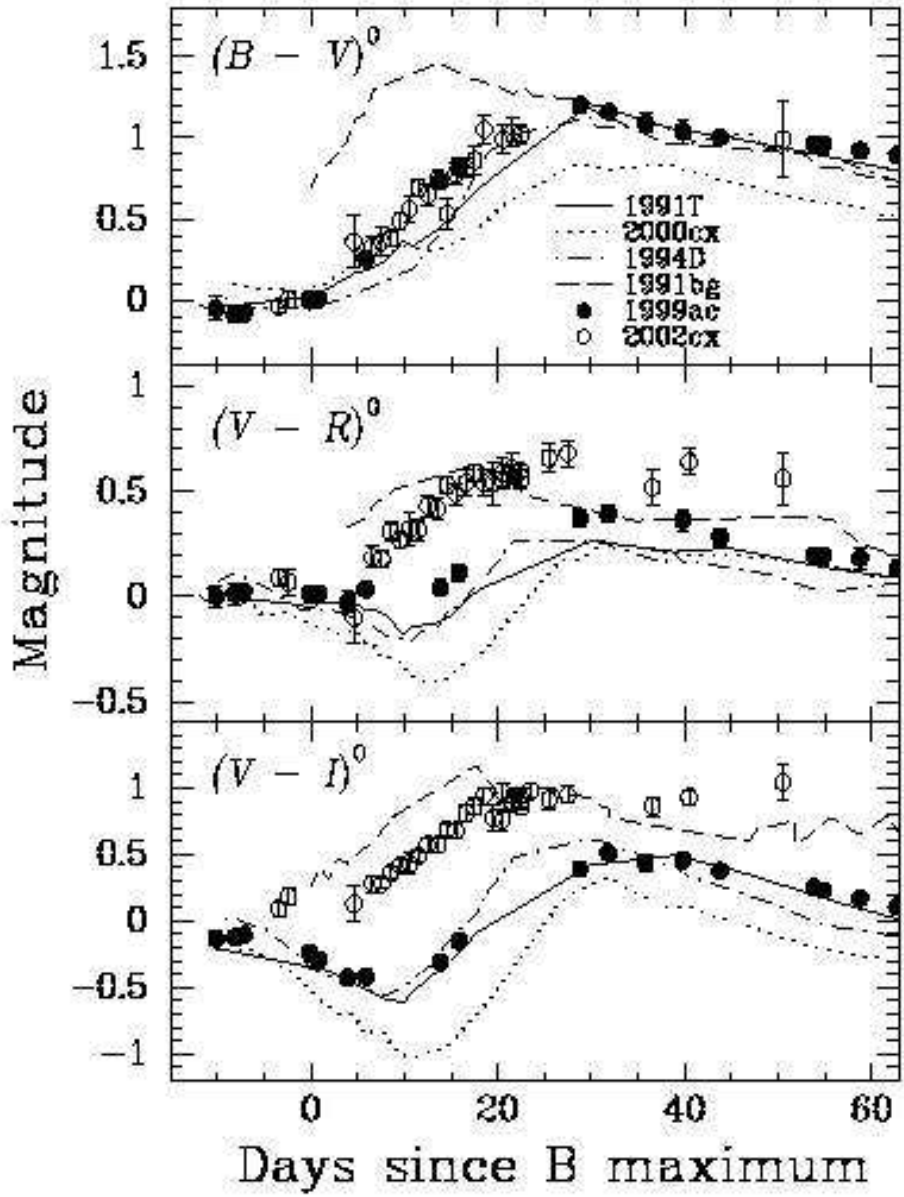}{6.4in}{-90}{100}{100}{-150}{560}}
\caption{Comparison between the intrinsic color evolution of SN 2002cx and
other SNe~Ia.}
\label{4}
\end{figure}

\begin{figure}
{\plotfiddle{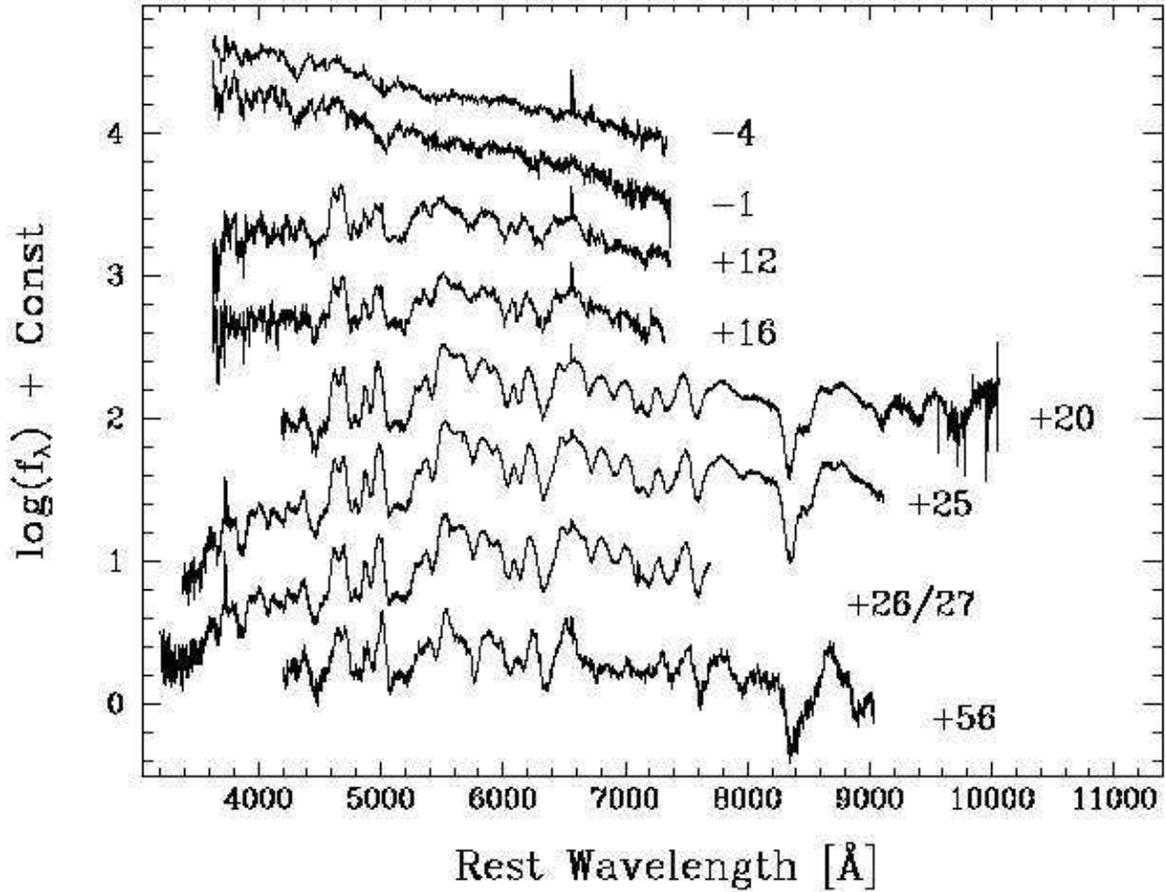}{6.4in}{-90}{80}{80}{-70}{560}}
\caption{Montage of spectra of SN 2002cx. The phases marked are relative
to the date of $B$ maximum. To improve clarity, the spectra have been 
shifted vertically by arbitrary amounts. The spectra have been corrected
for the redshift of the host galaxy ($cz = 7184$ km s$^{-1}$). The first four spectra
have been boxcar smoothed (smoothing box = 5 pixels). The Keck spectra on
days +26 and +27 have been combined (see text for details).}
\label{5}
\end{figure}

\begin{figure}
{\plotfiddle{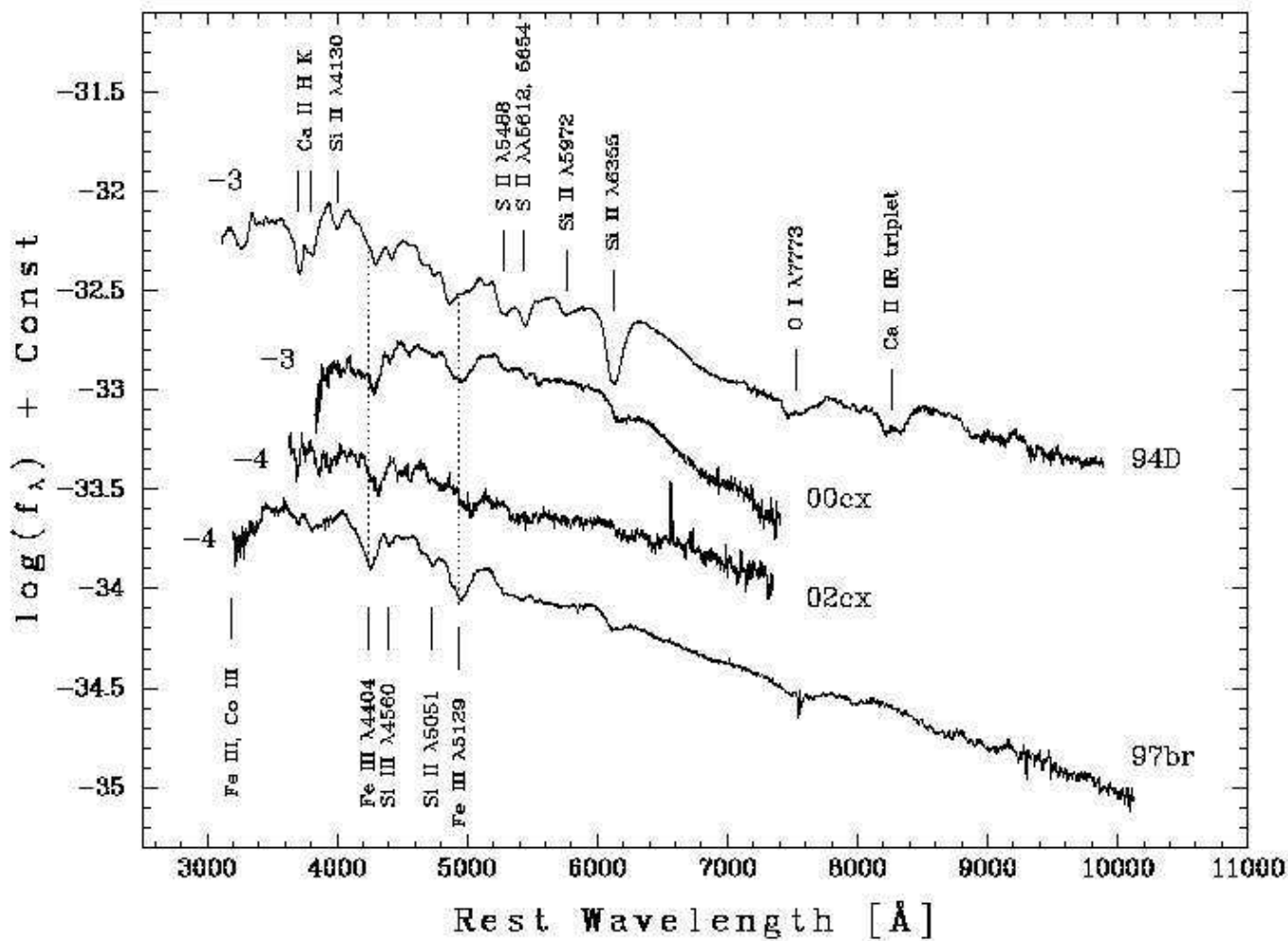}{6.4in}{-90}{80}{80}{-70}{560}}
\caption{The spectrum of SN 2002cx at $t$ = $-4$~d, shown with 
comparable-phase spectra of SNe 1994D, 2000cx, and 1997br. See
text for sources of spectra and line identifications. 
(Note that the continuum shape of SN 2000cx is incorrectly shown at this epoch,
due to problems with the spectrograph; Li et al. 2001b.)
The two dotted
vertical lines mark the position of the absorption minima for 
the Fe~III $\lambda$4404 and Fe~III $\lambda$5129 lines observed
in SN 1997br. Note the apparent shift of the SN 2002cx
absorption lines toward redder wavelengths.}
\label{6}
\end{figure}

\begin{figure}
{\plotfiddle{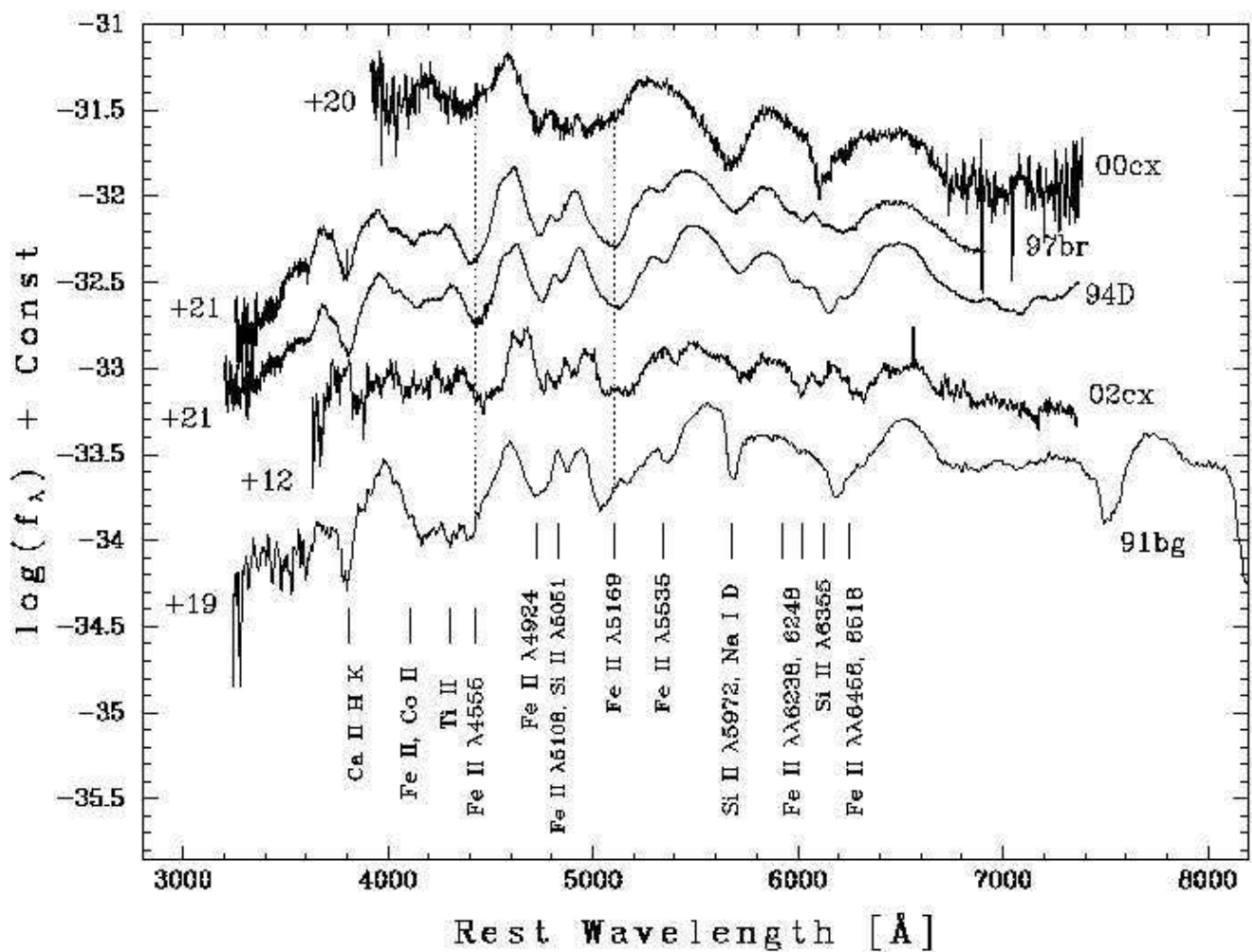}{6.4in}{-90}{80}{80}{-70}{560}}
\caption{The spectrum of SN 2002cx at $t$ = +12~d, shown with spectra
of other SNe~Ia at much older ages. The two dotted vertical lines  mark 
the position of the absorption minima for the Fe~II $\lambda$4555 and 
Fe~II $\lambda$5169 lines observed in SN 1994D. Note the strong Fe~II
lines observed in SN 2002cx, and the relatively weak Fe~II lines
in SN 2000cx. }
\label{7}
\end{figure}

\begin{figure}
{\plotfiddle{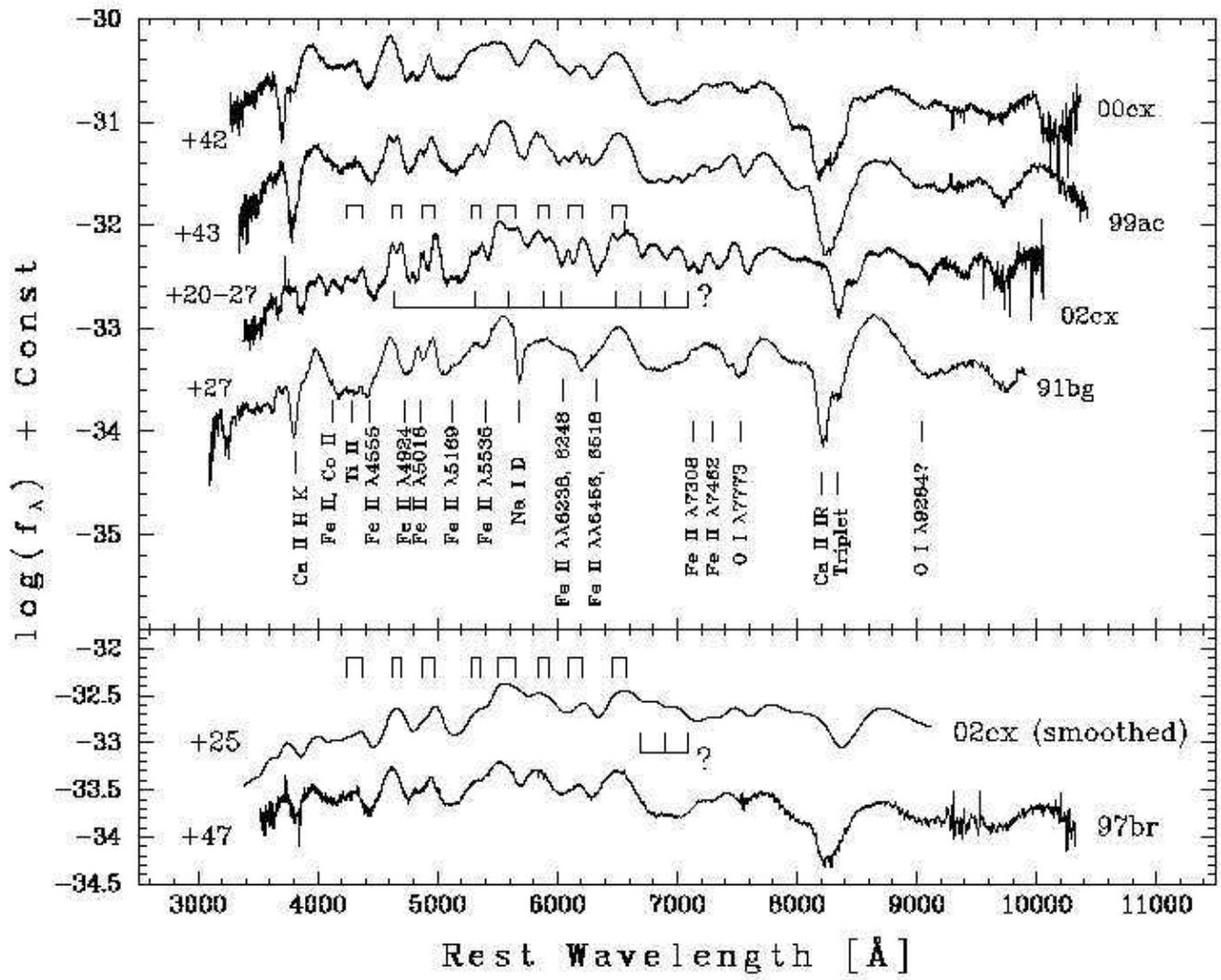}{6.4in}{-90}{80}{80}{-70}{560}}
\caption{The spectrum of SN 2002cx at $t$ = +20/25~d, shown with 
spectra of other SNe~Ia at older ages. The upper panel shows the line 
identifications and the comparison of the spectra. The pairs of short
vertical lines above the SN 2002cx spectrum mark possible ``double peaks,"
while these below the SN 2002cx spectrum mark possible additional resolved
lines (compared with other SNe~Ia). 
The lower panel shows the comparison between the $t$ = +25~d
spectrum of SN 2002cx after convolving with a Gaussian function with
$\sigma$ = 2,500 km s$^{-1}$, and the day +47 spectrum of SN 1997br.
Note that although the ``double peaks" are gone, additional features seem to be
present around 7000~\AA\ in the spectrum of SN 2002cx.}
\label{8}
\end{figure}

\begin{figure}
{\plotfiddle{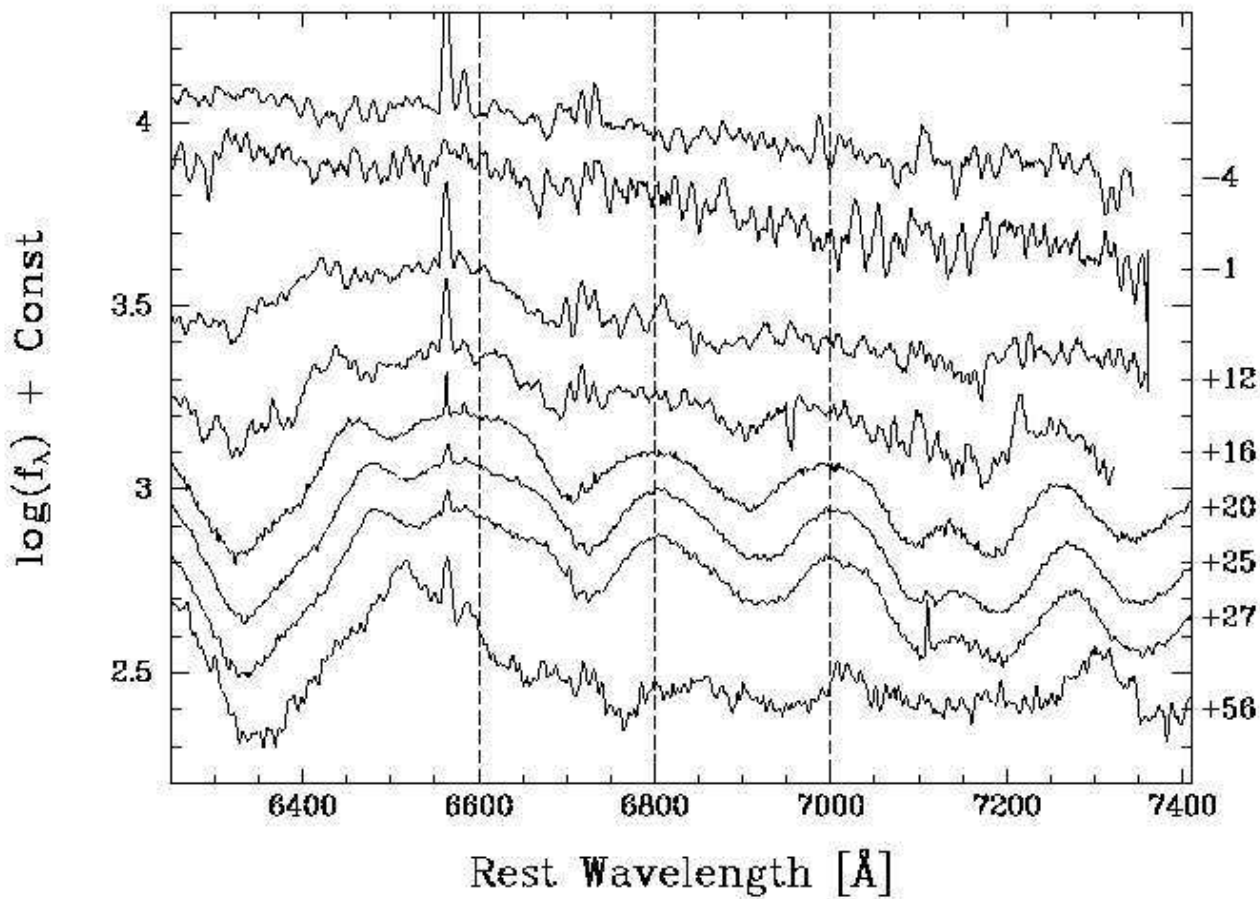}{6.4in}{-90}{80}{80}{-70}{560}}
\caption{Spectral evolution of SN 2002cx from 6200~\AA\ to 7400~\AA. 
The three dashed lines mark the approximate positions of the additional
emission lines seen in the convolved spectrum of Fig. 8. 
Note the comparable strength of these lines from $t$ = 
+20 to +27~d, and their disappearance or weakness in the $t$ = +56 spectrum.}
\label{9}
\end{figure}

\begin{figure}
{\plotfiddle{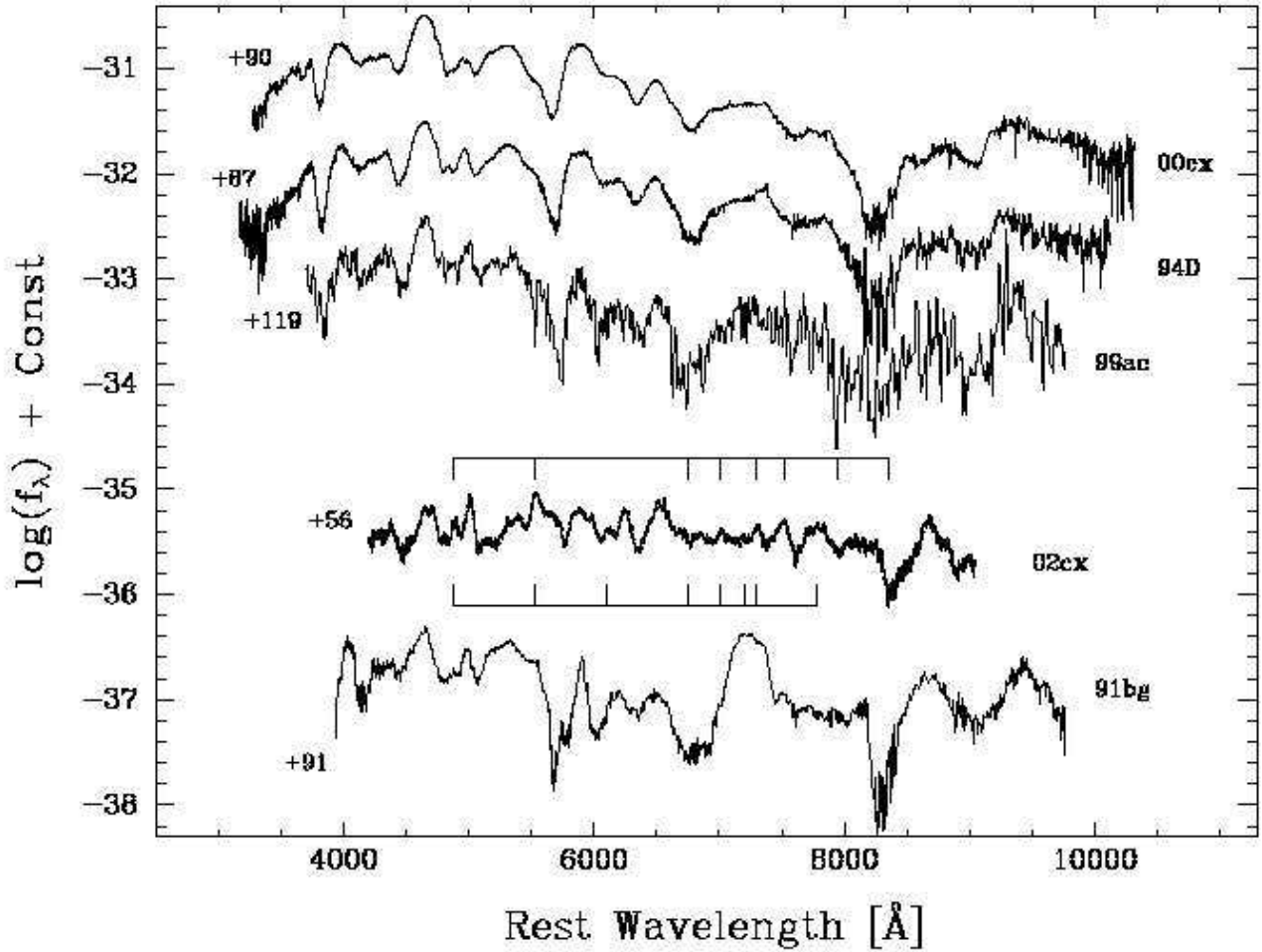}{6.4in}{-90}{80}{80}{-70}{560}}
\caption{Comparison of the nebular spectrum of SN 2002cx with those of other
SNe~Ia. The short vertical lines above the SN 2002cx spectrum mark the differences
from the spectra of SNe 2000cx, 1994D, and 1999ac, while lines below the
spectrum mark the differences from SN 1991bg. Note the dramatic differences
in the range 6500--8500~\AA.}
\label{10}
\end{figure}

\begin{figure}
{\plotfiddle{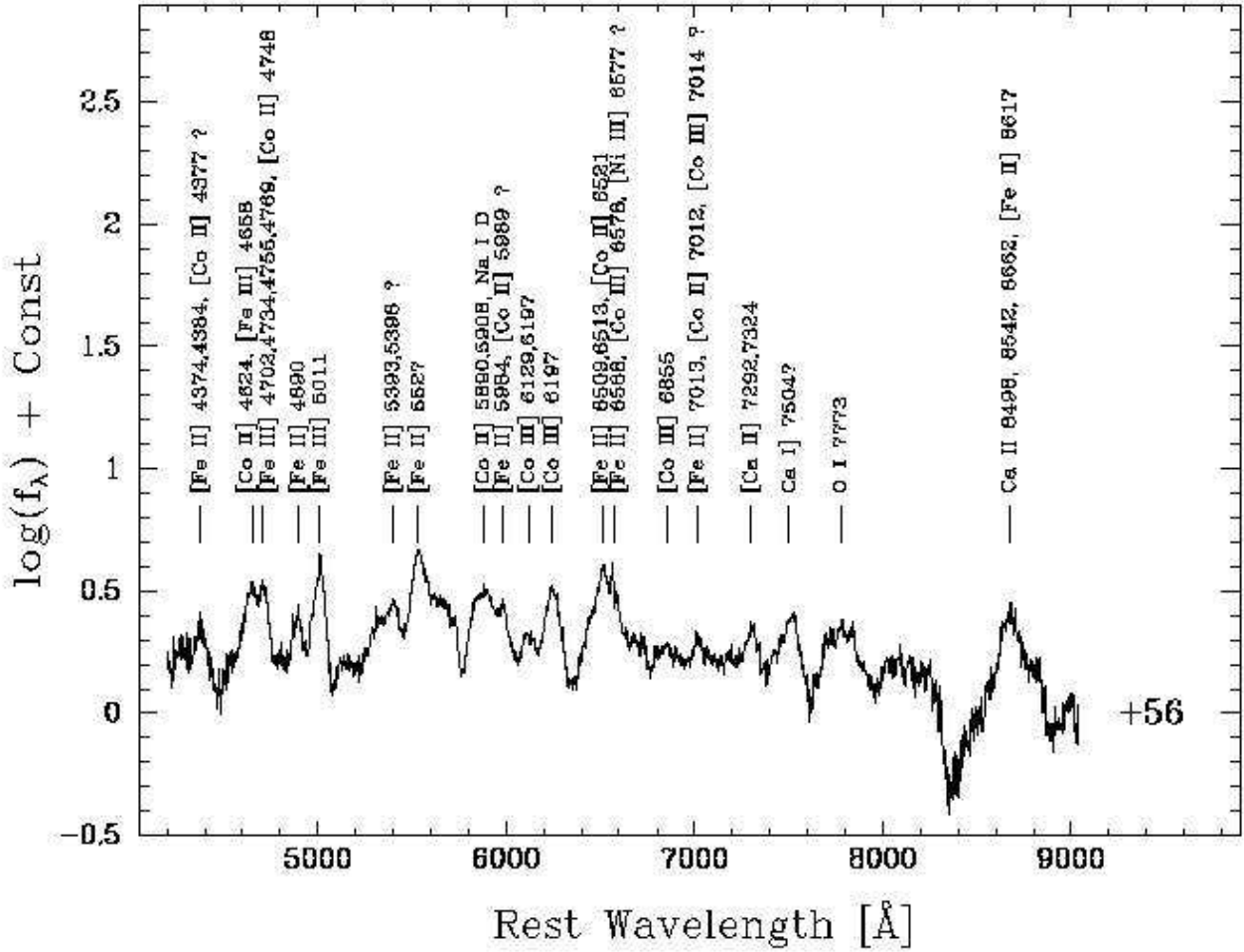}{6.4in}{-90}{80}{80}{-70}{560}}
\caption{Possible identifications of the lines in the nebular spectrum
of SN 2002cx. Most of the identifications are taken from Mazzali et al. (1997),
but those with a ``?" are based on wavelength coincidence with the [Fe~II] 
and [Co~II] lines identified in the Atomic Line List v2.04 (see text for 
details).}
\label{11}
\end{figure}

\begin{figure}
{\plotfiddle{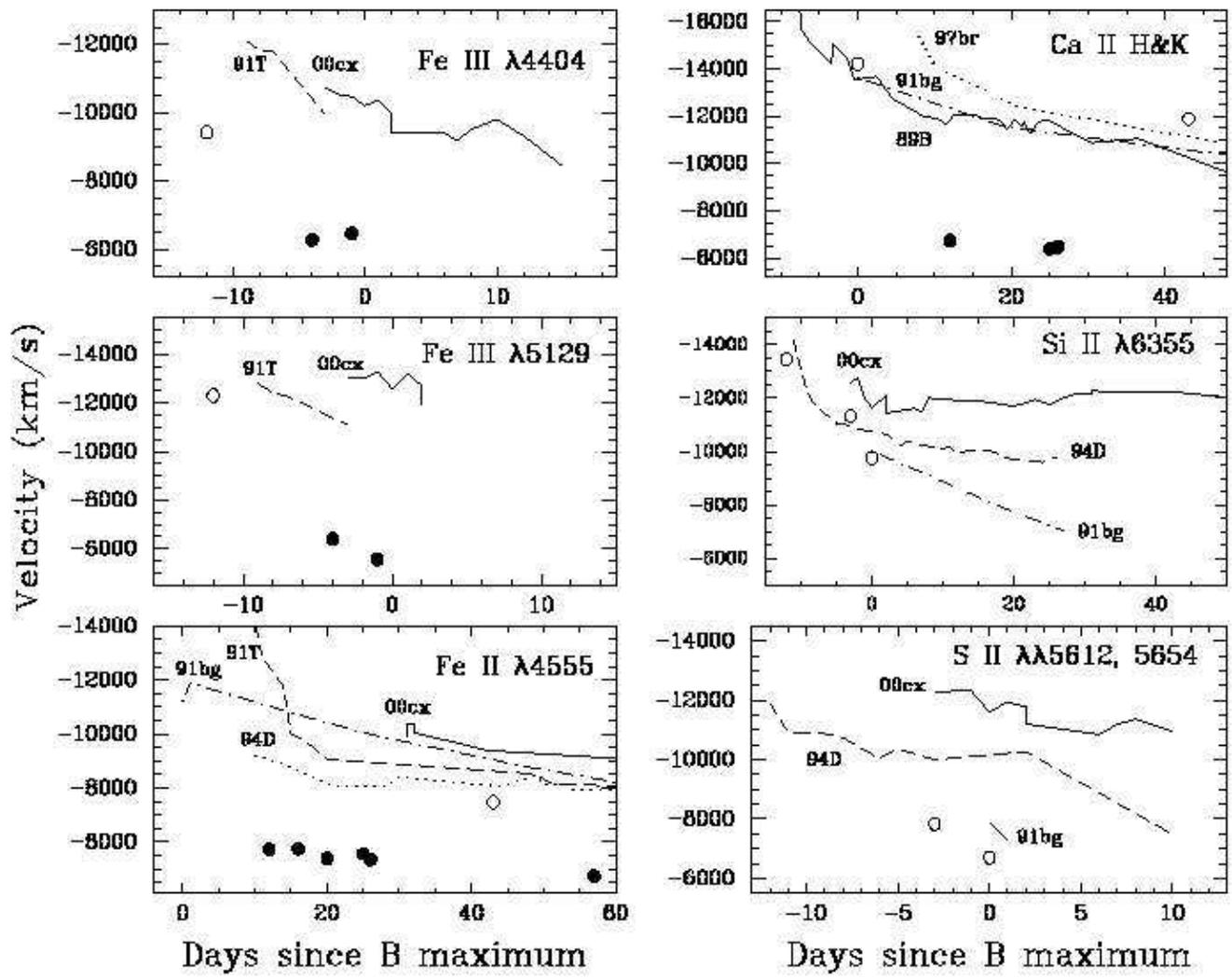}{6.4in}{-90}{80}{80}{-70}{560}}
\caption{Evolution of the expansion velocities as deduced from several 
absorption minima.  The measurements of SN 2002cx are shown with solid
circles, and those of SN 1999ac with open circles.  The expansion 
velocities of the other SNe are taken or measured from Wells et al.
(1994, for SN 1989B), Filippenko et al. (1992b, for SN 1991T), 
Leibundgut et al. (1993, for SN 1991bg), Filippenko (1997, for SN 1994D), 
Li et al. (1999, for SN 1997br), and Li et al. (2001b, for SN 2000cx).
Note the extremely low expansion velocities of SN 2002cx.}
\label{12}
\end{figure}

\newpage

\renewcommand{\arraystretch}{0.75}

\begin{deluxetable}{llllll}
\tablecaption{Photometry of comparison stars}
\tablehead{
\colhead{ID}&\colhead{$V$}&
\colhead{$(B-V)$}&\colhead{$(V-R)$}&
\colhead{$(V-I)$} & \colhead{$N_{calib}$}
}
\startdata
1 & 15.467(06) & 0.635(30) & 0.410(08) & 0.807(30) &2 \\
2 & 17.254(26) & 0.608(33) & 0.374(30) & 0.744(36) &3 \\
3 & 16.332(23) & 0.523(40) & 0.338(20) & 0.663(12) &2 \\
4 & 16.503(15) & 0.509(40) & 0.376(07) & 0.734(08) &2 \\
5 & 16.788(34) & 0.586(20) & 0.362(17) & 0.755(37) &6 \\
6 & 17.026(40) & 0.796(21) & 0.499(09) & 0.967(35) &6 \\
7 & 17.109(36) & 0.658(13) & 0.407(18) & 0.819(45) &6 \\
8 & 17.901(39) & 1.166(62) & 0.831(17) & 1.528(55) &6 \\
\enddata    
\tablenotetext{}{Note: All quantities are magnitudes. 
Uncertainties in the last two digits 
are indicated in parentheses.}
\end{deluxetable}

\begin{deluxetable}{cccccl}
\tablecaption{Photometry of SN 2002cx}
\tablehead{
\colhead{JD $-$} & \colhead{$B$} & \colhead{ $V$ } &
\colhead{$R$} & \colhead{$I$} &\colhead{Tel.}  \\
\colhead{2450000} & \colhead{(mag)} &
\colhead{(mag)} &
\colhead{(mag)} &
\colhead{(mag)} &
}
\startdata
2411.78 &17.81(03) &17.81(03) &17.70(02) &17.70(04) &Nickel\\
2412.78 &17.77(03) &17.73(04) &17.64(04) &17.53(04) &KAIT\\
2413.81 &17.69(04) &$-$ &$-$ &$-$ &KAIT\\
2419.81 &17.85(12) &17.45(10) &17.53(07) &17.30(08) &KAIT\\
2421.77 &18.06(05) &17.69(04) &17.48(03) &17.39(04) &KAIT\\
2422.75 &18.10(09) &17.70(03) &17.50(03) &17.40(03) &KAIT\\
2423.73 &18.22(03) &17.80(03) &17.47(02) &17.42(03) &KAIT\\
2424.75 &18.35(04) &17.82(03) &17.53(03) &17.38(03) &KAIT\\
2425.76 &18.47(06) &17.87(06) &17.53(05) &17.42(06) &KAIT\\
2426.72 &18.61(04) &17.88(03) &17.54(04) &17.37(03) &KAIT\\
2427.72 &18.68(05) &17.99(04) &17.54(03) &17.40(04) &KAIT\\
2428.76 &18.83(04) &18.05(04) &17.61(03) &17.46(04) &KAIT\\
2429.75 &18.73(10) &18.16(03) &17.61(03) &17.45(03) &KAIT\\
2430.71 &19.02(04) &18.20(04) &17.69(03) &17.50(03) &KAIT\\
2431.78 &19.12(04) &18.27(04) &17.71(04) &17.44(04) &KAIT\\
2432.69 &19.24(07) &18.34(03) &17.73(03) &17.46(05) &KAIT\\
2433.74 &19.45(08) &18.36(04) &17.80(04) &17.40(05) &Nickel\\
2434.70 &$-$ &18.31(07) &17.75(07) &17.52(07) &KAIT\\
2435.70 &$-$ &18.45(06) &17.84(04) &17.67(05) &KAIT\\
2435.75 &19.49(08) &18.46(03) &17.89(03) &17.47(06) &Nickel\\
2436.75 &$-$ &18.52(05) &17.89(05) &17.60(04) &KAIT\\
2436.77 &19.61(08) &18.54(03) &17.93(02) &17.64(03) &Nickel\\
2437.69 &$-$ &18.57(04) &17.99(03) &17.64(06) &KAIT\\
2437.70 &19.65(06) &18.59(02) &17.97(02) &17.71(02) &Nickel\\
2438.73 &$-$ &18.66(05) &$-$ &17.66(05) &KAIT\\
2440.69 &$-$ &18.75(06) &18.07(04) &17.82(05) &KAIT\\
2442.69 &$-$ &18.81(04) &18.11(04) &17.84(05) &KAIT\\
2443.73 &$-$ &$-$ &18.19(06) &17.90(06) &KAIT\\
2447.70 &$-$ &$-$ &18.30(08) &17.97(08) &KAIT\\
2451.74 &$-$ &18.94(07) &18.40(08) &18.06(05) &KAIT\\
2455.70 &$-$ &19.11(05) &18.45(04) &18.16(04)&KAIT\\
2456.70 &$-$ &$-$ &18.45(07) &18.17(04) &KAIT \\
2465.70 &20.36(21)&19.33(07) &18.75(06)&18.27(08)&Nickel \\
\enddata    
\tablenotetext{}{Note: Uncertainties in the last two digits 
are indicated in parentheses.}
\end{deluxetable}

\begin{deluxetable}{ccccc}
\tablecaption{Photometric information on SN 2002cx}
\tablehead{
\colhead{Filter} &\colhead{$B$} & \colhead{$V$} &
\colhead{$R$} & \colhead{$I$} }
\startdata

UT of max. &May 20.7 $\pm$ 1.0&May 23.0 $\pm$ 1.0&$-$&$-$ \\
Julian Date of max.&2452415.2 $\pm$ 1.0&2452417.5 $\pm$ 1.0&$-$&$-$ \\
Magnitude at max.& 17.68 $\pm$ 0.10&17.57 $\pm$ 0.15&17.50 $\pm$ 0.10&17.40 $\pm$
0.10\\
$\Delta m_{15}$ (mag) &1.29 $\pm$ 0.11 &0.73 $\pm$ 0.16&$-$ & $-$
 \\
\enddata
\end{deluxetable}

\begin{deluxetable}{llllccll}
\tablecaption{Journal of spectroscopic observations of SN 2002cx.}
\tablehead{
\colhead{UT Date} & \colhead{$t$\tablenotemark{a}} &\colhead{Tel.\tablenotemark{
b}}
&\colhead{Range\tablenotemark{c}}& \colhead{Air.\tablenotemark{d}} &
\colhead{Slit} &\colhead{Exp.} &\colhead{Observer\tablenotemark{e}} \\
& (day) & &{~~~~~(\AA)}&&(arcsec)&{~~(s)~~}&
}
\startdata
 2002-05-17 &$-$4&F1&3600--7400 &1.1&3.0 & 2400 &  PB\\
 2002-05-20 &$-$1&F1&3600--7400 &1.1&3.0 & 1200 &  MC\\
 2002-06-02 &+12 &F1&3600--7400 &1.1&3.0 & 2400 &  MC\\
 2002-06-06 &+16 &F1&3720--7500 &1.1&3.0 & 2400 &  PB\\
 2002-06-10 &+20 &K2&4000--10300 &1.0&0.75 &1200&  WS, RS\\  
 2002-06-15 &+25 &K1&3260--9340 &1.0&1.0 & 900 &  EB, SK\\  
 2002-06-16 &+26 &K1&3200--5800\tablenotemark{f} &1.1&1.0 & 1800 &  GS1\\  
 2002-06-17 &+27 &K1&5470--9230\tablenotemark{g} &1.1&1.0 & 600 &  GS1\\  
 2002-07-16 &+56 &K1&4300--9260 &1.3&1.0 & 1200 &  CF,GS2\\  
\enddata
\tablenotetext{a}{Days since $B$ maximum brightness (JD = 2452415.2), rounded to
 the nearest day.}
\tablenotetext{b}{F1 = FLWO 1.5-m + FAST spectrograph; K2 = Keck II 10-m + ESI; K1 = Keck I 10-m + LRIS.}
\tablenotetext{c}{Observed wavelength range of spectrum. }
\tablenotetext{d}{Average airmass of observations.}
\tablenotetext{e}{CF = C. Fassnacht; EB = E. Berger; GS1 = G. Smith; GS2 = G. Squires; MC = M. Calkins; PB = P. Berlind; RS = R. Simcoe;  SK = S. Kulkarni; WS = W. Sargent.}
\tablenotetext{f}{This spectrum is combined with the +27~d spectrum and discussed
together.}
\tablenotetext{g}{Only the spectrum in the range 5470$-$7800 \AA\, is used in
this study; the region redward of 7800~\AA\, has second-order contamination.}
\end{deluxetable}

\begin{deluxetable}{ccccccc}
\tablecaption{Absolute peak magnitudes of several SNe Ia\tablenotemark{a}}
\tablehead{
\colhead{SN} & \colhead{$\Delta m_{15}(B)$} & \colhead{$M_B^{max}$} & \colhead{$M_V^{max}$} & \colhead{$M_R^{max}$} & \colhead{$M_I^{max}$} & \colhead{source, note} }
\startdata
1991T & 0.95(05)& $-19.56(23)$ & $-19.59(19)$ & $-19.50(16)$ & $-19.21(13)$ & 1,b \\
2000cx& 0.93(04)& $-19.32(45)$ & $-19.42(45)$ & $-19.25(45)$ & $-18.93(45)$ & 2,c \\
1999ac& 1.30(09)& $-18.98(39)$ & $-19.04(39)$ & $-19.08(40)$ & $-18.77(40)$ & 3,d \\
1994D & 1.31(08)&$-18.95(18)$ & $-18.90(16)$ & $-18.85(15)$ & $-18.68(14)$ & 4,e \\
2002cx& 1.29(11)\tablenotemark{f}&$-17.55(34)$ & $-17.62(35)$ & $-17.67(34)$ & $-17.75(34)$ & 5,d \\
1991bg& 1.94(08) & $-16.54(32)$ & $-17.28(31)$ & $-17.58(31)$ & $-17.68(31)$ & 6,e \\
\hline
standard& 1.10& $-19.32(08)$ & $-19.31(06)$ & $-$ & $-18.96(07)$ & 1,g \\
\enddata
\tablenotetext{a}{Uncertainties in the last two digits are indicated in parentheses.}
\tablenotetext{b}{The Cepheid distance of NGC 4527 (Gibson \& Stetson 2001) is 
adopted. See Saha et al. (2001) for another measurement
of the Cepheid distance based on the same {\it Hubble Space Telescope} dataset.}
\tablenotetext{c}{The results of Li et al. (2001b) are converted to $H_0$ = 72
km s$^{-1}$ Mpc$^{-1}$ as adopted in this paper. The error bars are also modified to
include the uncertainty in $H_0$.}
\tablenotetext{d}{The distance is measured from $v_{CMB}$ and $H_0$ = 72 km
 s$^{-1}$ Mpc$^{-1}$.}

\tablenotetext{e}{The distance is measured with the surface brightness
 fluctuations (SBF) method. Note that the SBF method yields $H_0 \approx$ 75 km
 s$^{-1}$ Mpc$^{-1}$ (e.g., Tonry et al. 2000), and we do not attempt to
 reconcile the small difference with our adopted value of $H_0$. See Ajhar et
 al. (2001) for a discussion of the reconciliation of the SBF and SN~Ia
 distance scales.}
\tablenotetext{f}{The final, future photometry using galaxy template subtraction 
may yield higher $\Delta m_{15}(B)$ $\approx$ 1.6 mag, see \S~2.2 for details.}
\tablenotetext{g}{Adopted from Gibson \& Stetson (2001). These are the weighted
mean of the corrected [to $\Delta m_{15}(B) = 1.1$ mag] peak absolute
magnitudes of nine Cepheid-calibrated SNe~Ia. Note that Gibson \& Stetson derive
$H_0$ = 73 km s$^{-1}$ Mpc$^{-1}$, and we do not attempt to reconcile the small 
difference with our adopted value of $H_0$.}
\tablerefs{(1) Gibson \& Stetson 2001; (2) Li et al. 2001b; (3) Li et al. in 
preparation; (4) Richmond et al. 1995; (5) this paper; (6) Turatto et al. 1996.}
\end{deluxetable}

\end{document}